\documentclass[reprint, superscriptaddress, amsmath, amssymb, aps, twocolumn]{revtex4-1}

\usepackage{hyperref}
\usepackage{graphicx}
\usepackage[caption = false]{subfig}
\usepackage{amsthm}

\usepackage{bm}
\usepackage{tensor}

\usepackage{diagbox}

\newcommand{\mybra}[1]{\left\langle{#1}\right|}
\newcommand{\myket}[1]{\left|{#1}\right\rangle}

\newcommand{\sket}[1]{|{#1}\rangle}

\newcommand{\ketbra}[2]{\left|{#1}\middle\rangle\middle\langle{#2}\right|}
\newcommand{\ketsysbra}[3]{\left|{#1}\middle\rangle_{#2}\middle\langle{#3}\right|}

\newcommand{\sbraopket}[3]{\langle{#1}|{#2}|{#3}\rangle}
\newcommand{\prtl}[2]{\frac{\partial{#1}}{\partial{#2}}}
\newcommand{\Mod}[1]{\ (\mathrm{mod}\ {#1})}

\newcommand{\floor}[1]{\lfloor #1 \rfloor}

\newtheorem{thm}{Theorem}
\newtheorem{lmm}{Lemma}

\begin{document}

\title{Implementing arbitrary quantum operations via quantum walks on a cycle graph}

\author{Jia-Yi Lin}
    \affiliation{National Lab of Solid State Microstructure, Collaborative Innovation Center of Advanced Microstructures, and Department of Physics, Nanjing University, Nanjing 210093, China.}

\author{Xin-Yu Li}
    \affiliation{Institute for Brain Sciences and Kuang Yaming Honors School, Nanjing University, Nanjing 210023, China.}

\author{Yu-Hao Shao}
    \affiliation{Institute for Brain Sciences and Kuang Yaming Honors School, Nanjing University, Nanjing 210023, China.}

\author{Wei Wang}
    \email{wangwei@nju.edu.cn}
    \affiliation{National Lab of Solid State Microstructure, Collaborative Innovation Center of Advanced Microstructures, and Department of Physics, Nanjing University, Nanjing 210093, China.}
    \affiliation{Institute for Brain Sciences and Kuang Yaming Honors School, Nanjing University, Nanjing 210023, China.}

\author{Shengjun Wu}
    \email{sjwu@nju.edu.cn}
    \affiliation{National Lab of Solid State Microstructure, Collaborative Innovation Center of Advanced Microstructures, and Department of Physics, Nanjing University, Nanjing 210093, China.}
    \affiliation{Institute for Brain Sciences and Kuang Yaming Honors School, Nanjing University, Nanjing 210023, China.}
    \affiliation{Hefei National Laboratory, University of Science and Technology of China, Hefei 230088, China}

\date{\today}

\begin{abstract}
The quantum circuit model is the most commonly used model for implementing quantum computers and quantum neural networks whose essential tasks are to realize certain unitary operations.
The circuit model usually implements a desired unitary operation $U(N)$ by a sequence of single-qubit and two-qubit unitary gates from a universal set. Although this certainly facilitates
the experimentalists as they only need to prepare several different kinds of universal gates, the number of gates required to implement an arbitrary desired unitary operation is usually large. Hence the efficiency
in terms of the circuit depth or running time is not guaranteed.
Here we propose an alternative approach; we use a simple discrete-time quantum walk (DTQW) on a cycle graph to model an arbitrary unitary operation $U(N)$ without the need to decompose it into a sequence of gates of smaller sizes. Our model is essentially a quantum neural network based on DTQW.
Firstly, it is universal as we show that any unitary operation $U(N)$ can be realized via an appropriate choice of coin operators.
Secondly, our DTQW-based neural network can be updated efficiently via a learning algorithm, i.e., a modified stochastic gradient descent algorithm adapted to our network.
By training this network, one can promisingly find approximations to arbitrary desired unitary operations. With an additional measurement on the output, the DTQW-based neural network can also implement general measurements described by positive-operator-valued measures (POVMs). We show its capacity in implementing arbitrary 2-outcome POVM measurements via numeric simulation.
We further demonstrate that the network can be simplified and can overcome device noises during the training so that it becomes more friendly for laboratory implementations.
Our work shows the capability of the DTQW-based neural network in quantum computation and its potential in laboratory implementations.
\end{abstract}


\maketitle

\section{introduction}\label{sec:intr}
    Quantum walk \cite{Aharonov}, the quantum counterpart of classical random walk, has been widely applied in achieving various quantum information processing tasks \cite{Portugal}. Because of its quadratic enhancement of variances, the quantum walk plays a vital role in many quantum search algorithms and provides possible exponential speedups due to the quantum interference during the walk \cite{Kempe}.
    Moreover, various experimental implementations of quantum walks prove its feasibility in real-life circumstances of quantum information processing \cite{Kia}.

    On the other hand, machine learning is a core technology in the age of artificial intelligence. Since machine learning faces the challenge of the lack of computational power and quantum computing has a vast computational potential, the possibility of combining quantum computing and machine learning has been considered. Quantum neural networks (QNNs), a newer class of models in the field of quantum machine learning, operate on quantum computers and perform calculations using quantum effects like superposition, entanglement, and interference. Investigations on QNNs \cite{farhi2018classification,zhao2019building,mitarai2018quantum,dallaire2018quantum,amin2018quantum,CZoufal,VDunjko,MSchuld} have revealed their potential advantages, such as training and processing speedups.
    Despite significant developments in the growing field of quantum machine learning, the trade-offs between quantum and classical models have not been systematically studied. In particular, the question of whether quantum neural networks are more powerful than classical neural networks is still open \cite{SAaronson2015nph}.

    A gate-model QNN is a QNN constructed on a gate-model quantum computer using a sequence of unitaries with associated gate parameters \cite{farhi2018classification}.
    Recent developments, such as quantum generative adversarial networks and quantum circuit learning, have more general and diverse QNN structures \cite{dallaire2018quantum,mitarai2018quantum,gyongyosi2019training}.
    Researchers have already proved that typical quantum walks are universal for quantum computation \cite{childs2009universal,lovett2010universal,kurzynski2013quantum,bian2015realization,zhao2015experimental}.
    However, these works mainly focus on state processing, and many auxiliary systems should be employed in general.
    In contrast, what we are attempting to achieve in this work is a universal control of the quantum system to implement arbitrary quantum operations, without any auxiliary system.
    For this purpose, we shall introduce a QNN based on discrete-time quantum walks (DTQW) on a cycle graph with specifically parameterized coin operators.
    We choose the graph to be a cycle because it is simple for laboratory implementations.
    We will prove that the DTQW-based QNN is indeed capable of realizing arbitrary unitary evolution of the closed system.

    Determining the parameters of the DTQW-based QNN analytically is possible.
    However, any further adjustments on the network, such as a reduction in the number of circuit depth, will pose extraordinary difficulties for analytical methods. In contrast, we will show that such adjustments can be effectively made with gradient descent, a well-known optimization algorithm frequently employed to train machine learning models, including both classical and quantum neural networks \cite{darken1992learning,bengio2013advances}.
    Another significant advantage of using gradient descent is that explicitly decomposing the desired operator into a sequence of gates from a universal set is no longer necessary.
    Furthermore, we shall simplify the network in various ways to facilitate laboratory implementations.
    For example, we shall use only rotations along the x-axis as the gates involved in the DTQW.
    We can still find decent approximations of the desired quantum operations in this situation using our DTQW-based QNN.

    Our work is organized as follows. We first introduce our DTQW-based neural network in Sec.~\ref{sec:formalism} and then prove its universality for quantum control in Sec.~\ref{sec:universality}. We further modify gradient descent and apply it to our DTQW-based QNN in Sec.~\ref{sec:gradient}. Finally, we simplify the QNN in Sec.~\ref{sec:friendly} to facilitate the laboratory implementations.

\section{Quantum neural network based on discrete time quantum walk}\label{sec:formalism}
    The quantum neural network based on quantum gates, the gate-model QNN, was first introduced due to its high experimental feasibility \cite{farhi2018classification}.
    The gate-model QNNs utilize a series of unitary operations in a certain order to process the quantum state. The unitary operations involve adjustable parameters. By optimizing these parameters and encoding information to the input and output states, the gate-model QNNs are sufficient to solve various learning tasks.
    In this section, we introduce the DTQW on a cycle graph. We choose the graph to be a cycle because it is simple for laboratory implementation. We will show that such DTQW also involves a series of adjustable unitary gates and is sufficient to learn quantum operations. Thus the DTQW on a cycle graph can be treated as a special type of the gate-model QNN.

\begin{figure}[tbp]
        \includegraphics[width=0.80\columnwidth]{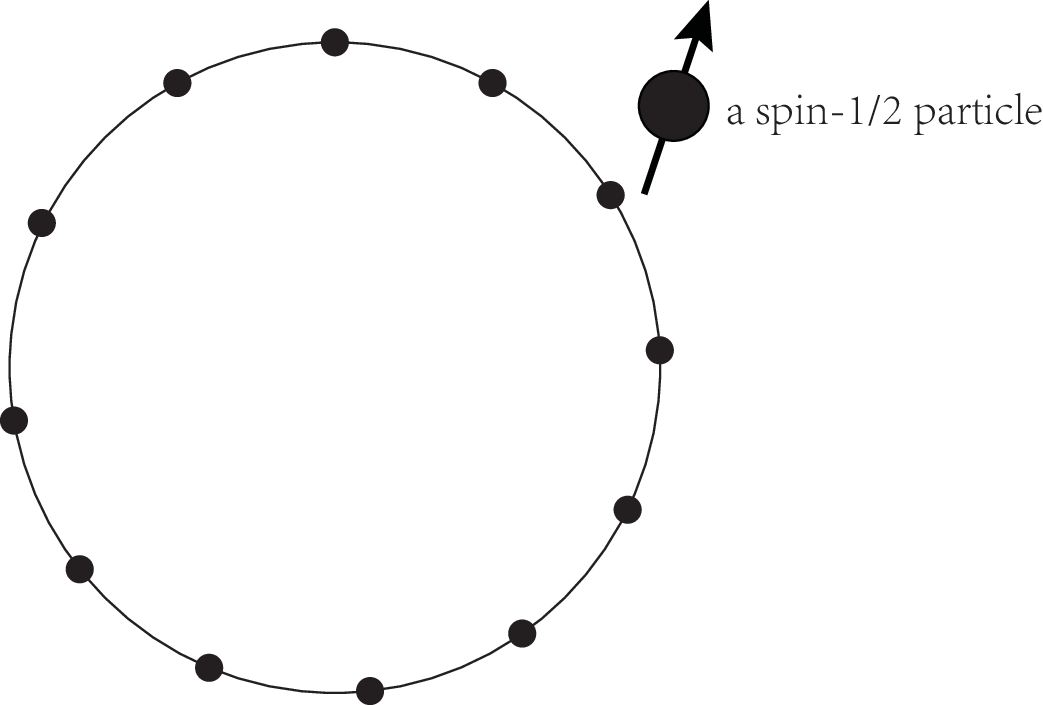}
        \caption{\label{fig:DTQW repr}
          This is a schematic representation of the DTQW on a cycle graph with 12 sites, i.e., $n=12$. The spin-1/2 particle represents the coin system of the DTQW. The dots on the cycle represent the positions that the walker possibly takes. The walker randomly walks on the sites for $T$-steps. The coin system and the position system combined becomes the total quantum system of the DTQW.
        }
    \end{figure}

    The DTQW on a cycle graph involves two Hilbert spaces,
        namely the coin space $\mathcal{H}_c$ and the position space $\mathcal{H}_p$,
        which are spanned by orthonormal basis $\{ \sket{0}_c, \sket{1}_c \}$
        and $\{ \sket{x}_p \}_{x=0}^{n-1}$ respectively,
        where $n$ is the number of sites in the cycle.
    The walker state $\sket{\Psi}$ is then in the space
        $\mathcal{H} = \mathcal{H}_c \otimes \mathcal{H}_p$.
    A schematic representation of the DTQW on a cycle graph is shown in Fig.~\ref{fig:DTQW repr}.
    The process of the DTQW is an iteration of
        applying coin operators $\hat{C}^{(t)}$
        and shift operators
        \begin{equation}\label{eq:shiftop}
            \hat{S} = \sum_{c=0}^1 \sum_{x=0}^{n-1}
                \ketbra{c,x+\delta_c \Mod{n}}{c,x}
        \end{equation}
        to the walker state, i.e.,
        \begin{equation}
            \sket{\Psi^{(t+1)}} = \hat{S} \hat{C}^{(t)} \sket{\Psi^{(t)}},
        \end{equation}
        where $t = 0,1,2,...$ denotes the ordinal of iterations,
        integer $\delta_c$ represents how far the walker is shifted
        if its coin is in the state $\sket{c}$.
    For simplicity, we choose $\delta_c = c$ throughout this letter.
    To make sure that the DTQW is flexible enough to implement various quantum operations, the coin operator $\hat{C}^{(t)}$ need to be site-dependent, i.e.,
        \begin{equation}\label{eq:Coin}
            \hat{C}^{(t)} = \sum_{x=0}^{n-1}
                \hat{c}_x^{(t)} \otimes \ketbra{x}{x},
        \end{equation}
        where $\hat{c}_x^{(t)} \in \mathrm{U}(2)$ flips the coin of the walker during the $t$-th iteration if the walker is at the site $x$.
        Since the operators $\hat{c}_x^{(t)}$ are applied to the coin only if the walker is at certain sites $x$, they are called single-site coin operators.


    Since the operations during every iteration are unitary,
        the total effect of a $T$-step DTQW
        \begin{equation}\label{eq:DTQW U}
            \hat{U}_{T,0} = \mathcal{T}\prod_{t=0}^{T-1} \hat{S}\hat{C}^{(t)}
        \end{equation}
        is also unitary,
        where $\mathcal{T}\prod$ denotes the time-ordered product.
    We define $\hat{U}_{t_1,t_0} = \mathcal{T}\prod_{t=t_0}^{t_1-1} \hat{S}\hat{C}^{(t)}$
        so that it is the time evolution operator, i.e.,
        $\sket{\Psi^{(t_1)}} = \hat{U}_{t_1,t_0} \sket{\Psi^{(t_0)}}$.
    One can notice that our version of the DTQW on a cycle graph
        is a straightforward generalization of the conventional Hadamard walk
        of which $\hat{c}_x^{(t)}=\hat{H}$ and $\delta_c=1-2c$.

\begin{figure}[tbp]
        \centering
            \includegraphics[width=0.95\columnwidth]{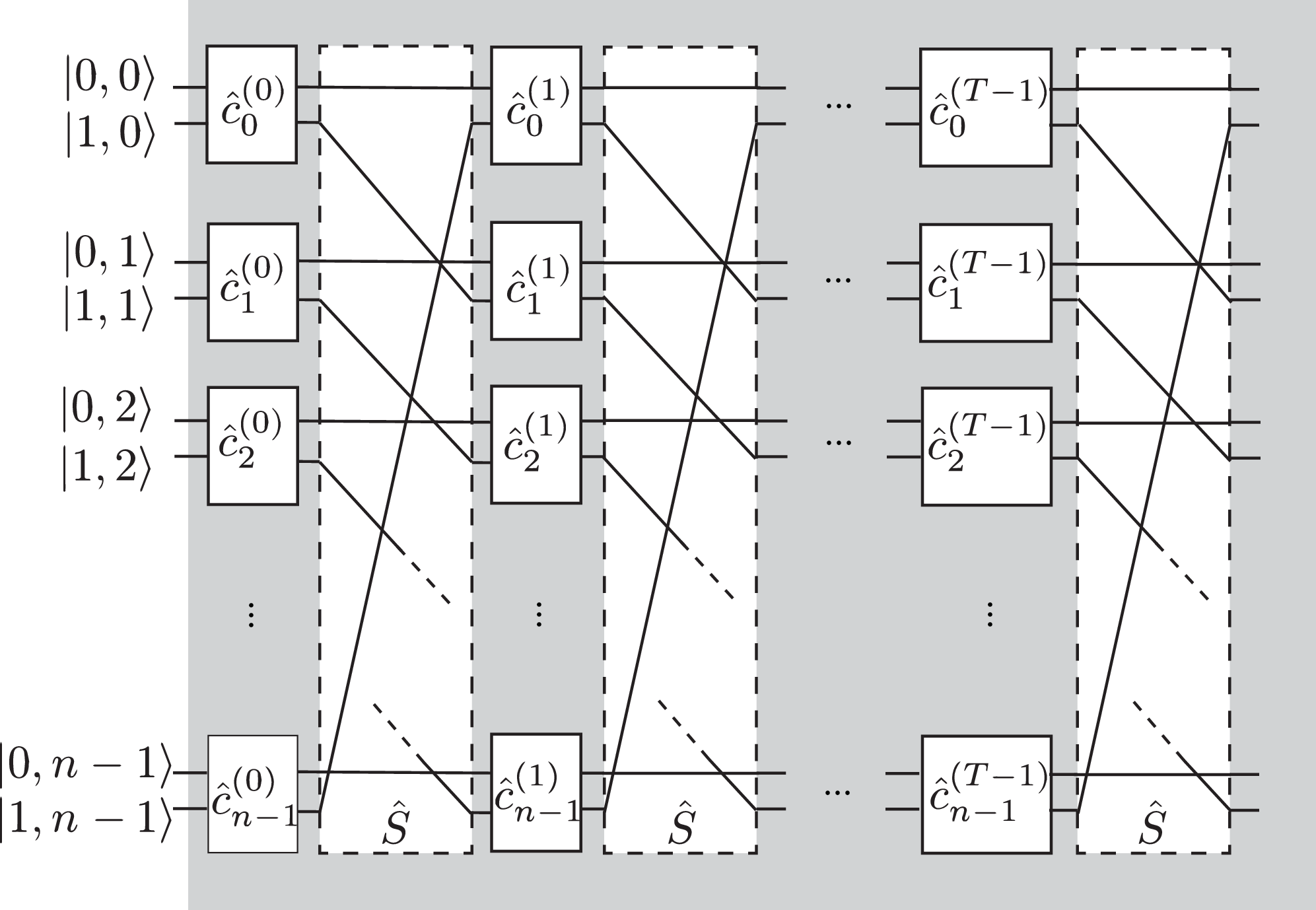}
        \caption{\label{fig:QNN repr}
            The DTQW on a cycle graph is represented in a fashion similar to gate-model QNNs. The operators $\hat{c}_x^{(t)}$ in little boxes are single-site coin operators of the DTQW, while the operators $\hat{S}$ in large dashed boxes are shift operators. Each $\hat{c}_x^{(t)}$ acts on two energy levels. These $\hat{c}_x^{(t)}$ are adjusted so that the total effect of the DTQW meets one's needs.
        }
    \end{figure}

    Every step of DTQW is unitary and is parameterized by $\hat{c}_x^{(t)} \in \mathrm{U}(2)$. These operators $\hat{c}_x^{(t)}$ can be treated as the adjustable gates in a gate-model quantum neural network. By adjusting these gates $\hat{c}_x^{(t)}$, we can use the DTQW to implement various quantum operations. Therefore the DTQW can be seen as a special type of gate-model quantum neural network.
    A schematic representations of the quantum neural network based on the DTQW on a cycle graph is shown in Fig.~\ref{fig:QNN repr}.
    The circuit depth of this network is the number of walking steps $T$ of the DTQW.
    In this work, we will denote the quantum neural network based on the DTQW on a cycle graph simply as the DTQW-QNN. We call the system of the quantum walker the underlying system of the DTQW-QNN.

\section{Universality and complexity of DTQW-based neural network}\label{sec:universality}

   The implementation of quantum operations via the DTQW on a cycle graph is one of the primary motivations of our work.
   The universality of DTQW for quantum computation has been shown in general \cite{childs2009universal,lovett2010universal}.
   While the previous work mainly focuses on the mapping from the initial state to the final state in a certain small subspace of the total system, in this work we take the overall effect on the total system into account.
    In this section, we investigate the capacity and universality of the DTQW on a cycle graph in implementing quantum operations,
    and show that it is universal for unitary operations, which is the main theorem of this section.

    By saying that the DTQW on a cycle graph is universal, we mean that
        any unitary operation on the overall Hilbert space
        $\mathcal{H} = \mathcal{H}_c \otimes \mathcal{H}_p$
        can be realized by a DTQW.
    Hence it is not only universal for computation but also universal for controlling the whole quantum system.
    To be more formal and specific, the following theorem is provided.

    \begin{thm}\label{thm:universality}
        For any unitary operator $\hat{V}\in\mathrm{U}(2n)$,
            there exists a positive integer $T$
            and a family of single-site coin operators
            $\{\hat{c}_x^{(t)}\}\subset\mathrm{U}(2)$
            indexed by the set
            $\{(x,t): 0\leq x<n \mbox{ and } 0\leq t<T\}$
            such that
            the total effect of the $T$-step DTQW is $\hat{V}$, i.e.,
            $\hat{U}_{T,0} = \hat{V}$,
            as long as $\delta_0 \neq \delta_1$
            and $\gcd(|\delta_0 - \delta_1|, n)=1$.
    \end{thm}

    We prove the universality of the DTQW on a cycle by decomposing arbitrary unitary operators $\hat{V}$ into a product of two-level unitary operators $\hat{V} = \hat{u}_m, \dots, \hat{u}_2 \hat{u}_1$ and construct a DTQW to implement every $\hat{u}_i$ for $i=1,2,\dots,m$. A detailed proof is provided in Appendix~\ref{app:verif}.

    As a demonstration of Theorem~\ref{thm:universality}, we first implement the controlled NOT (CNOT) gate with a DTQW on a cycle with two sites.
    We can find that according to Eq.~(\ref{eq:shiftop}), the shift operator
    \begin{equation}
        \hat{S} = \begin{bmatrix}
                            1 & 0& 0& 0\\
                            0 & 1& 0& 0\\
                            0 & 0& 0& 1\\
                            0 & 0& 1& 0
                        \end{bmatrix}
    \end{equation}
    is just the CNOT gate we need.
    Hence a simple one-step DTQW is equivalent to the CNOT gate if we choose all the single-site coin operators $\hat{c}_{x}^{(t)}$ to be the identity operator.

    Next, let us consider a more complicated two-level unitary operator, a unitary $\hat{U}$ controlled by two qubits
    \begin{equation}\hat{V}
        =
        \begin{bmatrix}
            1 & 0& 0& 0 & 0 & 0& 0& 0\\
            0 & 1& 0& 0 & 0 & 0& 0& 0\\
            0 & 0& 1& 0 & 0 & 0& 0& 0\\
            0 & 0& 0& 1 & 0 & 0& 0& 0\\
            0 & 0& 0& 0 & 1 & 0& 0& 0\\
            0 & 0& 0& 0 & 0 & 1& 0& 0\\
            0 & 0& 0& 0 & 0 & 0& a& b\\
            0 & 0& 0& 0 & 0 & 0& c& d
        \end{bmatrix},
    \end{equation}
    where $a,b,c,d$ are four matrix elements of $\hat{U}$.
    Comparing this operator $\hat{V}$ with the general form of two-level operators in Eq.~(\ref{eq:2-level V}), we can find that $c_0 = c_1 = 1$, $x_0 = 3$ and $x_1 = 4$.
    By substituting $c_0, c_1, x_0, x_1$ in Eqs.(\ref{eq:sub1}) and (\ref{eq:sub2}) with their respective values, we get \begin{equation}\label{eq:ccu}
        \hat{c}_x^{(t)} =
        \begin{cases}
            \hat\sigma_x & \mbox{if $t=0$ or $4$, and $x=4$} \\
            \begin{bmatrix}
                d& c\\
                b& a
            \end{bmatrix} & \mbox{if $t=1$ and $x=4$} \\
            \hat{I}_c & \mbox{otherwise}
        \end{cases}
    \end{equation}
    where $\hat\sigma_x$ is the Pauli $x$ matrix.
    By choosing the single-site coin operators $\hat{c}_x^{(t)}$ according to Eq.(\ref{eq:ccu}), we can realize the unitary operator $\hat{V}$ with an eight-step DTQW on a cycle with four sites.

    For the most general two-level unitary operators $\hat{V}$, the calculation is essentially the same as the above example, i.e., find the values of $c_0, c_1, x_0, x_1$ by comparing $\hat{V}$ with Eq.~(\ref{eq:2-level V}) and then substitute them in Eqs.~(\ref{eq:sub11}) and (\ref{eq:sub22}) if $c_0 = c_1$ or Eqs.~(\ref{eq:sub1}) and (\ref{eq:sub2}) if otherwise.
    For unitary operators $\hat{V}$ which are not two-level, we decompose them into a product of two-level unitary operators $\hat{V} = \hat{u}_m, \dots, \hat{u}_2 \hat{u}_1$ \cite{Nielsen2007Quantum}.
    By combining the DTQWs for $\hat{u}_i$ one after one, we can realize $\hat{V}$ with the final combined DTQW.
    As an example, the calculation to implement the Fourier transformation is provided in Appendix~\ref{app:fourier}.


    Implementing a unitary operation with the construction
        in the proof of Theorem~\ref{thm:universality} as above
        involves numerous steps of the walk.
    To reduce the number of steps, we provide
        in Appendix~\ref{app:optim}
        a further optimized scheme for implementations.
        With this scheme, no more than $2n^2-2n+1$ steps of walk is needed for the DTQW-QNN to be universal,
        where $n$ is the number of sites in the cycle.

\section{\label{sec:gradient} finding approximations via gradient descent}
    It is sometimes cumbersome
        to find exact realizations of desired quantum operations in analytical ways.
    However, fair approximations to desired operations
        are often acceptable for practical purposes.
    In this section,
        we introduce an algorithm in a machine learning fashion
        to find the approximations
        by applying gradient descent to the DTQW-QNN.
    With this algorithm, the required number of depth can be further reduced
        when approximations are allowed.

        In order to apply gradient descent to the DTQW-QNN,
            we have to do the following three things in advance.
        \begin{enumerate}
            \item Parameterize the single-site coin operators
                with a four dimensional real vector $\vec\alpha^{(x,t)}$:
                \begin{equation}\label{eq:site coin general}
                    \hat{c}_{x}^{(t)} =
                        e^{i \alpha_{3}^{(x,t)} \hat\sigma_{3}}
                        e^{i \alpha_{2}^{(x,t)} \hat\sigma_{2}}
                        e^{i \alpha_{1}^{(x,t)} \hat\sigma_{1}}
                        e^{i \alpha_{0}^{(x,t)} \hat\sigma_{0}} ,
                \end{equation}
                in which $\hat\sigma_{j}$ is the $j$th Pauli matrix,
                    $\hat\sigma_{0}=\hat{I}$.
            \item Introduce a state-wise loss function $L_{\sket{\Psi}}$:
                \begin{equation}\label{eq:loss}
                    L_{\sket{\Psi}} = \frac{1}{2} \left\lVert
                        \sket{\Psi^{(T)}} - \sket{\Phi^{(T)}}
                    \right\rVert^2 ,
                \end{equation}
                where $\sket{\Psi^{(T)}} = \hat{U}_{T,0}\sket{\Psi}$ and
                    $\sket{\Phi^{(T)}} = \hat{V}\sket{\Psi}$
                    are the final state and the desired final state respectively.
            \item Derive the partial derivative:
                \begin{equation}\label{eq:partial}
                    \prtl{L_{\sket{\Psi}}}{\alpha^{(x,t)}_j} =
                        \operatorname{Im}\left(
                            \sbraopket{\Phi^{(t)}}{
                                \hat\Sigma_j^{(x,t)}
                            }{\Psi^{(t)}}
                        \right),
                \end{equation}
                where $\sket{\Psi^{(t)}}=\hat{U}_{t,0}\sket{\Psi}$ and
                    $\sket{\Phi^{(t)}}=\hat{U}_{T,t}^\dagger\sket{\Phi^{(T)}}$
                    are the forward-propagation and back-propagation states respectively,
                $\hat\Sigma_j^{(x,t)} = (
                    \hat{n}_j^{(x,t)}\cdot{\vec\sigma}
                )\otimes\ketbra{x}{x} + \sum_{\xi\neq x}\hat{I}_c\otimes\ketbra{\xi}{\xi}$,
                    ${\vec\sigma}=\sum_{j=0}^3\hat\sigma_j \vec{e}_j$,
                    and $\hat{n}_0^{(x,t)}$,
                    $\hat{n}_1^{(x,t)}$,
                    $\hat{n}_2^{(x,t)}$,
                    $\hat{n}_3^{(x,t)}$
                    equals
                    \begin{equation*}
                        \begin{bmatrix}
                            1\\0\\0\\0
                        \end{bmatrix},
                        \begin{bmatrix}
                            0\\1\\0\\0
                        \end{bmatrix},
                        \begin{bmatrix}
                            0\\0\\
                            \cos{2\alpha_1^{(x,t)}}\\
                            \sin{2\alpha_1^{(x,t)}}
                        \end{bmatrix},
                        \begin{bmatrix}
                            0\\
                            \sin{2\alpha_2^{(x,t)}}\\
                            -\cos{2\alpha_2^{(x,t)}}\sin{2\alpha_1^{(x,t)}}\\
                            \cos{2\alpha_2^{(x,t)}}\cos{2\alpha_1^{(x,t)}}
                        \end{bmatrix}
                    \end{equation*}
                    respectively.
        \end{enumerate}

        Gradient descent iteratively moves the parameters
            in the opposite direction of the gradient, i.e.,
            \begin{equation}\label{eq:update}
                \mbox{new }\alpha_j^{(x,t)} \leftarrow
                    \mbox{old }\alpha_j^{(x,t)}
                    - \eta \prtl{L_{\sket{\Psi}}}{\alpha_j^{(x,t)}} ,
            \end{equation}
            where $\eta$ is a positive real number called learning rate.
        Hence, the loss gradually drops during the iteration
            and the approximation to $\hat{V}$ by $\hat{U}_{T,0}$ becomes better and better.

        The details of the algorithm to find the parameters of the DTQW-QNN
            $\left\{ \vec\alpha^{(x,t)} : 0\leq x<n \mbox{ and } 0\leq t<T\right\}$
            to approximate a desired unitary operator $\hat{V}$
            are as the following.
        \begin{enumerate}
            \item Set the total number of depth $T$
                and the learning rate $\eta$
                to be an appropriate positive integer and real respectively.
            \item Randomly initialize all the parameters $\alpha^{(x,t)}_j$.
            \item \label{it:begin} Randomly sample a state $\sket{\Psi}$
                from the total Hilbert space $\mathcal{H}$.
            \item Calculate the partial derivatives
                $\prtl{L_{\sket{\Psi}}}{\alpha^{(x,t)}_j}$
                for all $t$, $x$, $j$
                according to Eq.~(\ref{eq:partial}).
            \item \label{it:end} Update all the parameters
                according to Eq.~(\ref{eq:update}).
            \item Repeat Steps~\ref{it:begin}~to~\ref{it:end}
                until an acceptable approximation is reached.
        \end{enumerate}

        \begin{figure}[tbp]
            \centering
            \includegraphics[width=0.9\columnwidth]{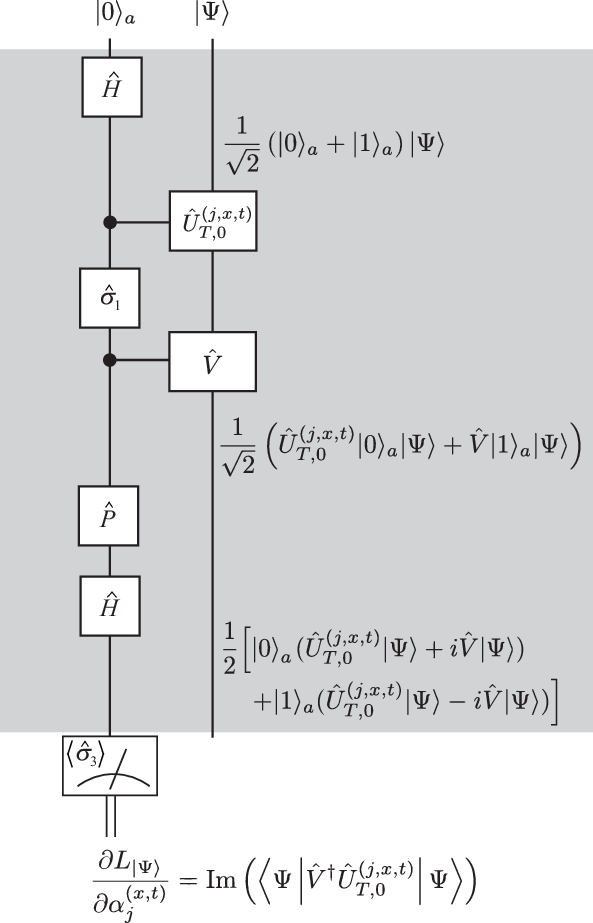}
            \caption{\label{fig:measure gradients}
                The circuit to calculate the gradient with a measurement on an ancillary qubit.
                The order of the operators applied is from top to bottom.
                The operator $\hat{U}_{T,0}^{(j,x,t)}$ equals $\hat{U}_{T,t}\hat{\Sigma}_j^{(x,t)}\hat{U}_{t,0}$.
                Some states during the computation with this circuit are listed on the right side.
                At the last, the average value $\langle\hat{\sigma}_3\rangle$ of the ancillary qubit is measured.
            }
        \end{figure}
        One can notice that our choice of the loss function
            leads to a friendly form of gradients Eq.~(\ref{eq:partial})
            for numerical calculation.
        The states $\myket{\Psi^{(t)}}$ and $\myket{\Phi^{(t)}}$
            can be calculated by a forward-propagation and a back-propagation efficiently.
        Moreover, the gradients can be calculated by implementing
            a circuit with the help of an ancillary qubit as shown in Fig.~\ref{fig:measure gradients}.
        At the last of the circuit, the average value $\langle\hat{\sigma}_3\rangle$ of the ancillary qubit is measured.
        The result $\langle\hat{\sigma}_3\rangle$ can be used to update the parameters of the DTQW-QNN since $\langle\hat{\sigma}_3\rangle$ always coincides with the partial derivative $\partial{L_{\sket{\Psi}}} / \partial{\alpha^{(x,t)}_j}$ in Eq.~(\ref{eq:partial}).
        This might enable us to implement
            simultaneous tomography and cloning of an unknown unitary operation.

        Besides, the position space $\mathcal{H}_p$ is commonly much larger
        than the coin space $\mathcal{H}_c$.
        Theorem~\ref{thm:universality} thus indicates that
        one can indirectly control a large system by
        controlling a small two-level coin system
        via DTQW on a cycle graph.
        For example, unitary operations and general two-outcome measurements described by positive-operator-valued measures (POVMs)
        can be applied to the position space in this way
        straightforwardly according to Theorem~\ref{thm:universality}.
        If we are only interested in the unitary operators that act on the position space $\mathcal{H}_p$,
        we only need one arbitrary site to be allowed to assign nonidentity coin operators.
        The detailed content is provided in Appendix~\ref{app:ctrl large sys}.

        \subsection*{Numerical results}

\begin{figure}[htbp]
            \centering
            \includegraphics[width=0.5\textwidth]{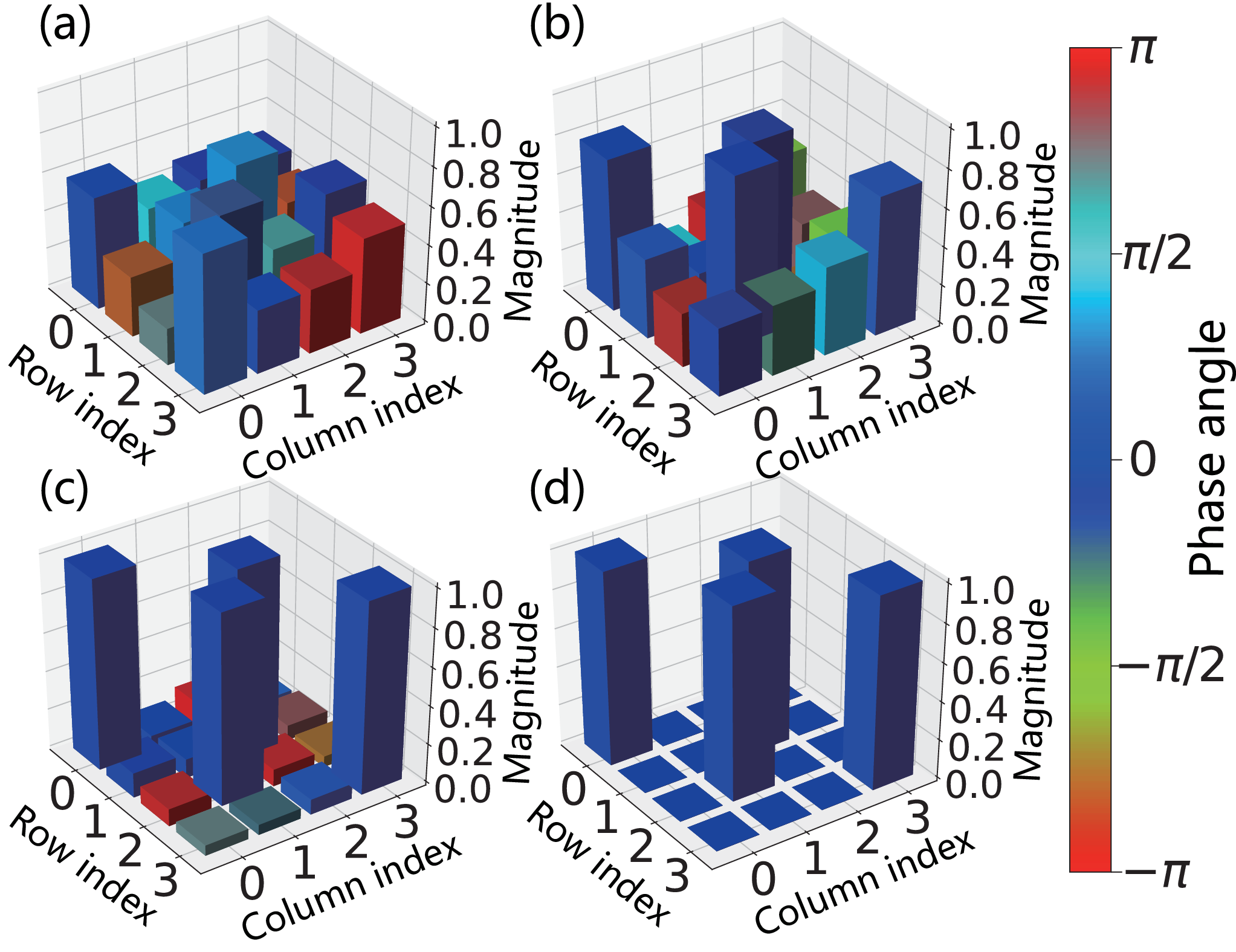}
            \caption{
                The training of a SWAP gate is visualized; each subfigure shows the matrix elements of the unitary operator $\hat{U}_{T,0}$ represented by the DTQW-QNN in different stages of the training:
                (a) shows the random matrix picked up by the DTQW-QNN before the training, and (b-d) give the updated matrix after $30$, $60$, and $240$ updates to the parameters of the DTQW-QNN respectively. After $240$ updates, the DTQW-QNN represents a SWAP gate precisely.
                Each bar corresponds to a matrix element, which is a complex number, of the unitary operator $\hat{U}_{T,0}$.
                The labels on the bottom left and right corners represent, respectively, the row indices and the column indices of the matrix elements.
                The height of a bar represents the magnitude of the matrix element of $\hat{U}_{T,0}$, while the color of a bar represents its phase angle.
            }
            \label{fig:SWAP}
        \end{figure}

        We first test our algorithm with a DTQW-QNN to learn the $\mathrm{SWAP}$ gate.
        Because all matrix elements of $\mathrm{SWAP}$ are either $0$ or $1$,
        it would be visually clear whether a unitary operator is close to $\mathrm{SWAP}$ after the operator is visualized.
        The change of the DTQW unitary operator $\hat{U}_{T,0}$ during the training is visualized in Fig.~\ref{fig:SWAP}.
        As the DTQW-QNN is trained, $\hat{U}_{T,0}$ becomes closer and closer to the desired gate $\mathrm{SWAP}$.
        And the DTQW-QNN realizes the $\mathrm{SWAP}$ after the training is finished.

        To measure how well the DTQW $\hat{U}_{T,0}$ approximates the desired unitary $\hat{V}$, we introduce the distance \begin{equation}
            d(\hat{U}_{T,0},\hat{V})= \sqrt{
                1 - \left|\mathrm{tr}(\hat{U}_{T,0}\hat{V}^\dagger)/{2n}\right|^2
            } .
        \end{equation}
        between the operators $\hat{U}_{T,0}$ and $\hat{V}$.
        The smaller this distance is, the better the DTQW approximates the desired operator.
             \begin{figure}[tbp]
                \centering
                \includegraphics[width=0.95\columnwidth]{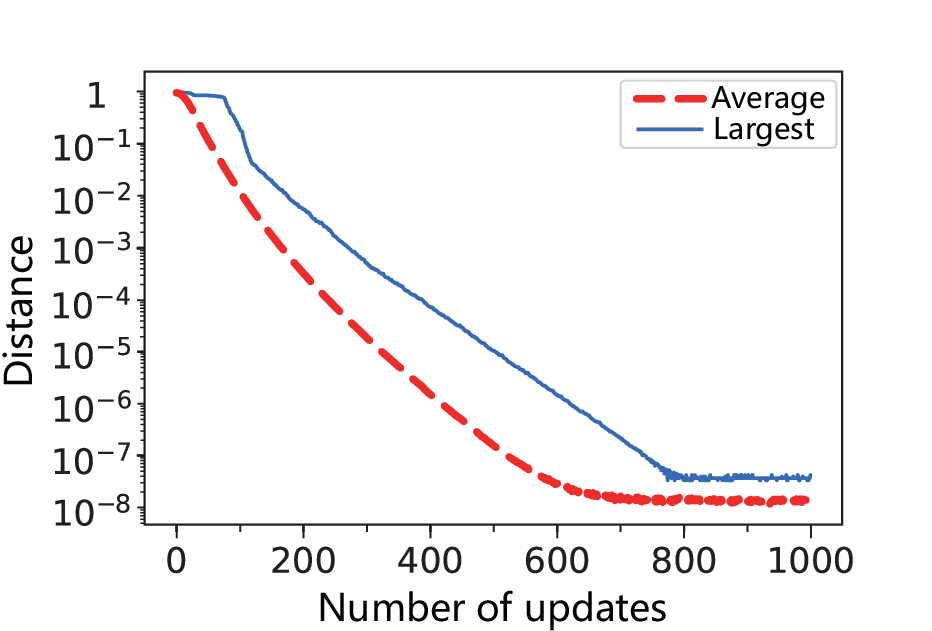}
                \caption{
                The evolution of the distance $d(\hat{U}_{T,0},\hat{V})$ between the DTQW unitary operator and the desired operator during the training of DTQW-QNNs.
                The horizontal axis represents the number of times (epochs) that the DTQW-QNN is updated. The vertical axis represents the distance.
                The thick dashed red line represents the average of distances calculated from 200 sampled operators $\hat{V}$ and their corresponding DTQW-QNNs, while the thin blue line represents the distance of the worst sample, i.e., the largest distance among those samples.}
                \label{fig:dist distri V total}
            \end{figure}

\begin{figure}[htbp]
                \centering                    \includegraphics[width=\columnwidth]{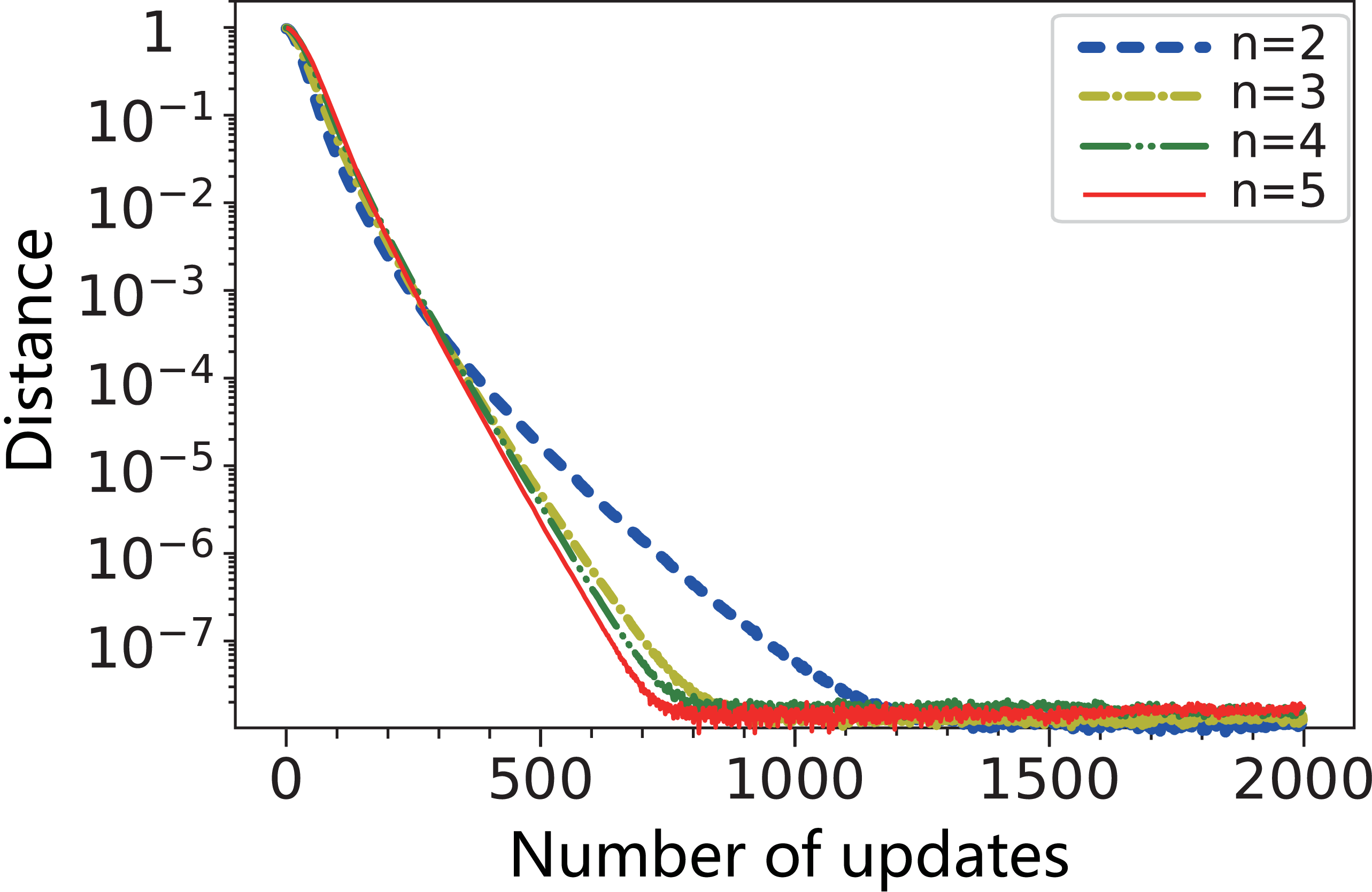}
                \caption{\label{fig:dist drop V total}
                    The distance $d(\hat{U}_{T,0},\hat{V})$ evolves differently during the training of DTQW-QNNs with different number $n$ of sites on the cycle.
                    The horizontal axis represents the number of times that the DTQW-QNN is updated. The vertical axis represents the distance $d(\hat{U}_{T,0},\hat{V})$.
                }
            \end{figure}

            In order to show that the DTQW-QNN can actually approximate arbitrary unitary operator,
            we sample 200 desired operators $\hat{V}$
                from $\mathrm{U}(4)$ according to the Haar measure
                and train 200 DTQW-QNNs in parallel to approximate these operators $\hat{V}$ respectively.
            The evolution of the distance during the training is plotted in Fig.~\ref{fig:dist distri V total}.
            After the training, the final distance between the DTQW-QNN and the desired operator is smaller than $10^{-7}$ even for the worst case of the 200 samples.
            For DTQW-QNNs with different number $n$ of sites on the cycle, Fig.~\ref{fig:dist drop V total} shows that the average distance is also always smaller than $10^{-7}$. From Fig.~\ref{fig:dist drop V total}, we can also notice that with more sites on the cycle, the training of the DTQW-QNNs is faster, i.e., less updates are needed.

\begin{figure}[tbp]
                \centering
                \includegraphics[width=\columnwidth]{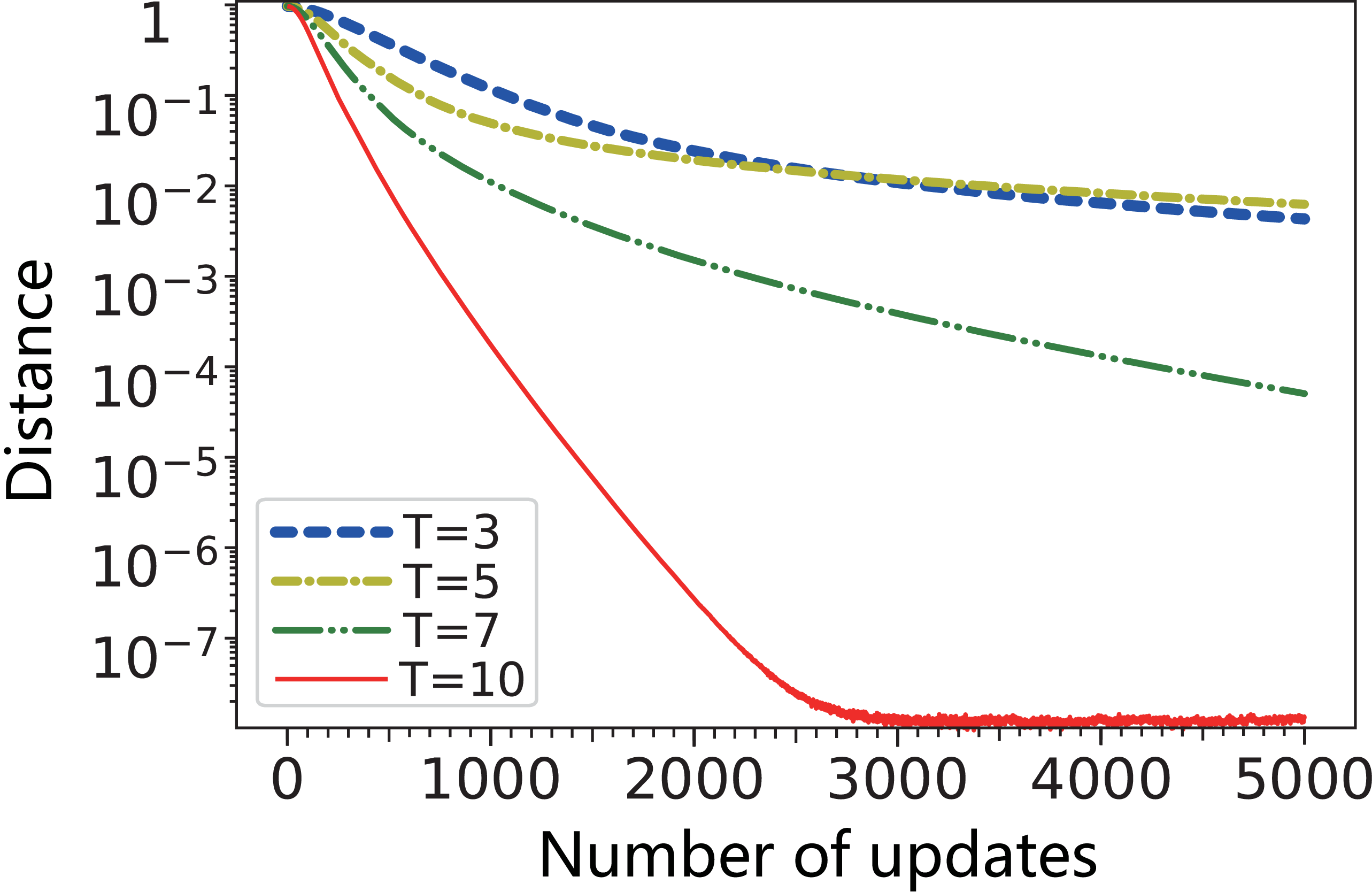}
                \caption{\label{fig:acc by overpara}
                    The distance $d(\hat{U}_{T,0},\hat{V})$ of DTQW-QNNs with a larger depths $T$ drops faster during the training.
                    The horizontal axis represents the number of times that the DTQW-QNN is updated. The vertical axis represents the distance $d(\hat{U}_{T,0},\hat{V})$.
                }
            \end{figure}

            Training the DTQW-QNN exhibits some similar phenomena as
                training classical machine learning models.
            For example, the implicit acceleration by overparameterization \cite{arora2018on}
                also emerges in the training of the DTQW-QNN.
            The implicit acceleration by overparameterization is a phenomenon where the neural network training becomes faster if more layers are added to the network.
            For DTQW-QNNs, more layers mean more steps of walk, i.e., a larger depth $T$.
            As shown in Fig.~\ref{fig:acc by overpara}, when the number of depth $T$ is larger, the distance drops faster during the training.

            \begin{figure}[tbp]
                \centering
                %
                \includegraphics[width=\columnwidth]{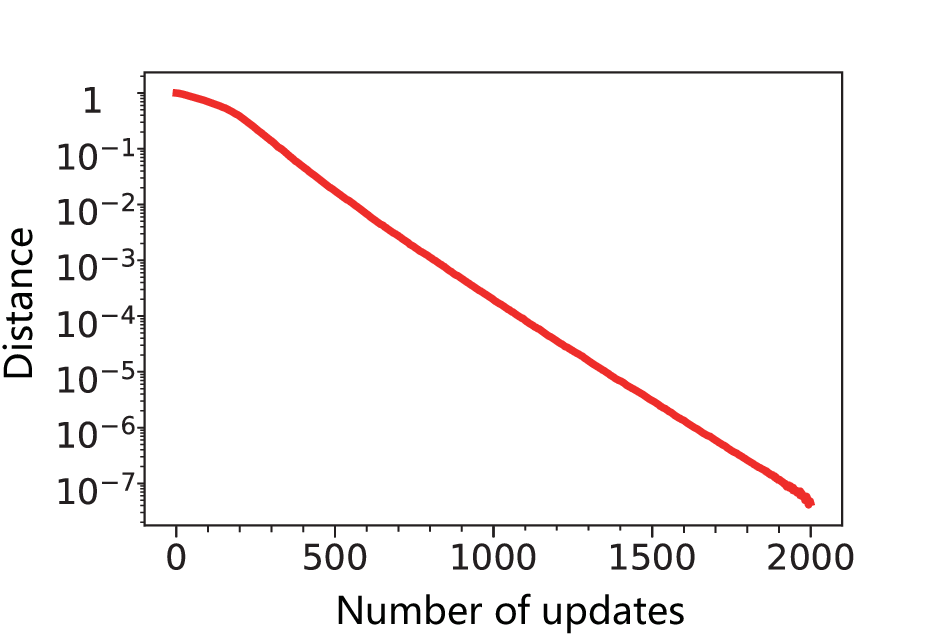}
                \caption{\label{fig:large sys}
                    The evolution of the distance $d(\hat{U}_{T,0},\hat{V})$ during the training of a DTQW-QNN with a $40$-dimensional underlying quantum system, i.e., there are $20$ sites on the cycle.
                    The horizontal axis represents the number of times that the DTQW-QNN is updated.
                    The vertical axis represents the distance.
                }
            \end{figure}

            To show that the algorithm also works for larger quantum systems,
                we apply it to a DTQW on a cycle graph with $20$ sites as a demonstration to realize the quantum Fourier transformation.
            As shown in Fig.~\ref{fig:large sys}, this DTQW-QNN with a $40$-dimensional underlying quantum system can still be trained to implement the operator we want.
            For the meta parameters used to generate the numerical results throughout this work, see Appendix~\ref{app:metapara}.

    \section{Making the DTQW-based neural network more friendly for implementations}\label{sec:friendly}

        In all previous parts of this work,
            we have assumed that the single-site coin operators $\hat{c}_x^{(t)}$
            can take values from $\mathrm{U}(2)$ arbitrarily.
        This means the single-site coin operator can have arbitrary phase and arbitrary rotational axis.
        However, it would be much easier to implement rotations along a fixed axis with fixed phases in laboratories.
        Hence, in this section, we simplify the DTQW so that it becomes easier to implement.
        Also, there are always noises when DTQW-QNNs are implemented in laboratories. We test it under the situation where noises are presented in the single-site coin operators $\hat{c}_x^{(t)}$.
        Throughout this section, the numerical demonstrations are all based on DTQWs on a cycle with two sites.

            \begin{figure}[tbp]
                \centering
                \subfloat[\label{fig:fixed random phases}]{
                    \centering
                    \includegraphics[width=\columnwidth]{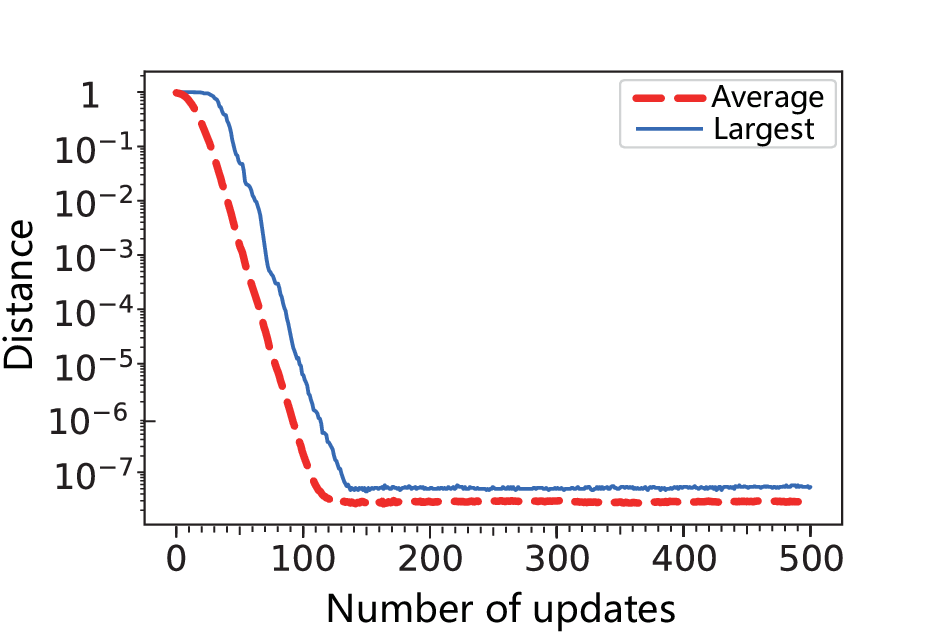}
                }\\
                \subfloat[\label{fig:fixed rotation axes}]{
                    \centering
                    \includegraphics[width=\columnwidth]{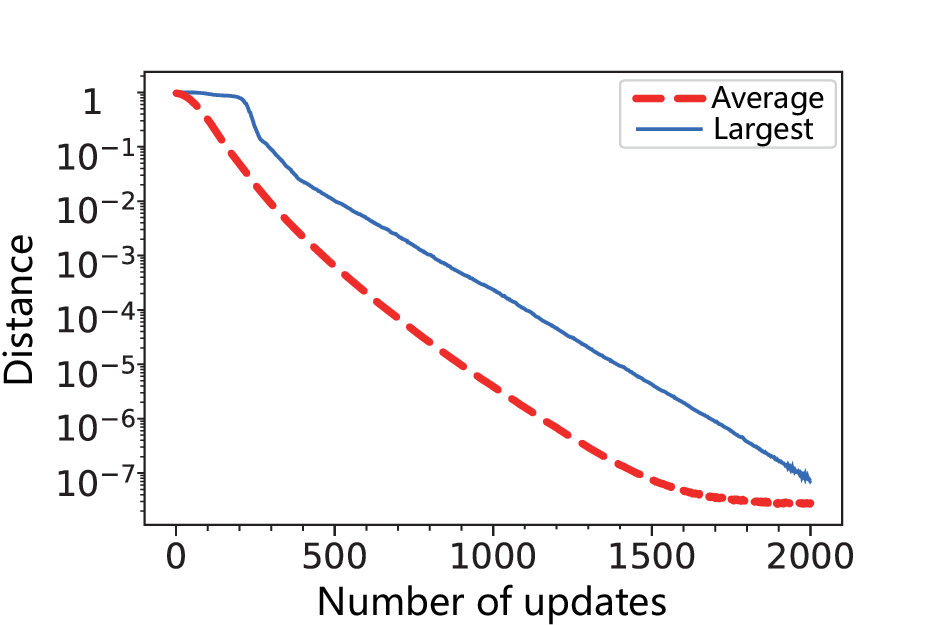}
                }\\
                \caption{\label{fig:abc}
                    The evolution of the distance $d(\hat{U}_{T,0},\hat{V})$ during the training of a DTQW-QNN with different simplifications:
                    (a) the phases of single-site coin operators are random and fixed;
                    (b) all single-site coin operators are rotations along the $x$-axis.
                        The horizontal axis represents the number of times that the DTQW-QNN is updated. The vertical axis represents the distance.
                        The thick dashed red line shows the average of distances calculated from 200 sampled operators $\hat{V}$ and their corresponding DTQW-QNNs, while the thin blue line shows the largest distance among those samples.
                }
            \end{figure}

        \subsection{Random fixed phases}\label{sec:fix rand phase}
            Firstly, it can be observed that
                the phases $e^{i \alpha_{0}^{(x,t)} \hat\sigma_{0}}$
                in Eq.~(\ref{eq:site coin general})
                of single-site coin operators $\hat{c}_x^{(t)}$
                are relative phases when $\hat{c}_x^{(t)}\otimes\ketbra{x}{x}$ are summed in Eq.~(\ref{eq:Coin}).
            They are not merely a contribution to the global phase of the DTQW $\hat{U}_{T,0}$.
            Hence, any change in one of the phases may cause
                a nontrivial change in $\hat{U}_{T,0}$.
            This seemingly requires an annoying tuning of all the phase factors of single-site coin operators at different times $t$ and at different sites $x$ when the DTQW is implemented.

            Fortunately, we find that these phase factors $e^{i \alpha_{0}^{(x,t)}\hat\sigma_{0}}$ actually need no adjustment.
            As shown in Fig.~\ref{fig:fixed random phases},
                the DTQW-QNN can still approximate an arbitrary operator $\hat{V}$ via gradient descent
                even if
                all the phase factors signed to different sites are random
                and fixed during the training, i.e.,
                \begin{equation}
                    \hat{c}_{x}^{(t)} =
                        e^{i\bm{a}^{(x)}}
                        e^{i \alpha_{3}^{(x,t)} \hat\sigma_{3}}
                        e^{i \alpha_{2}^{(x,t)} \hat\sigma_{2}}
                        e^{i \alpha_{1}^{(x,t)} \hat\sigma_{1}}
                       ,
                \end{equation}
                where phases $\bm{a}^{(x)}$ are independent real random variables.
            This releases us from the cumbersome tuning of the phases of single-site coin operators.


        \subsection{Simple rotations along x-axis only}\label{sec:fixed axis}
            The formalism of the single-site coin operators $\hat{c}_x^{(t)}$ in Eq.~(\ref{eq:site coin general}) involves three consecutive rotations,
                namely, $e^{i \alpha_{j}^{(x,t)} \hat\sigma_{j}}, j=1,2,3$,
                each along a different axis.
            To make it easier for laboratory implementations, we simplify the single-site coin operators to be simple rotations only along the $x$-axis, i.e.,
                \begin{equation}\label{eq:sraxo}
                    \hat{c}_{x}^{(t)} =
                        e^{i\bm{a}^{(x)}}
                        e^{i \alpha^{(x,t)} \hat\sigma_{1}} ,
                \end{equation}
                where $\alpha^{(x,t)}$ now is merely a real parameter.
            In this situation the DTQW-QNN can still realize arbitrary operators via gradient descent,
                as indicated by Fig.~\ref{fig:fixed rotation axes}.

\begin{figure}
                    \centering
                    \includegraphics[width=\columnwidth]{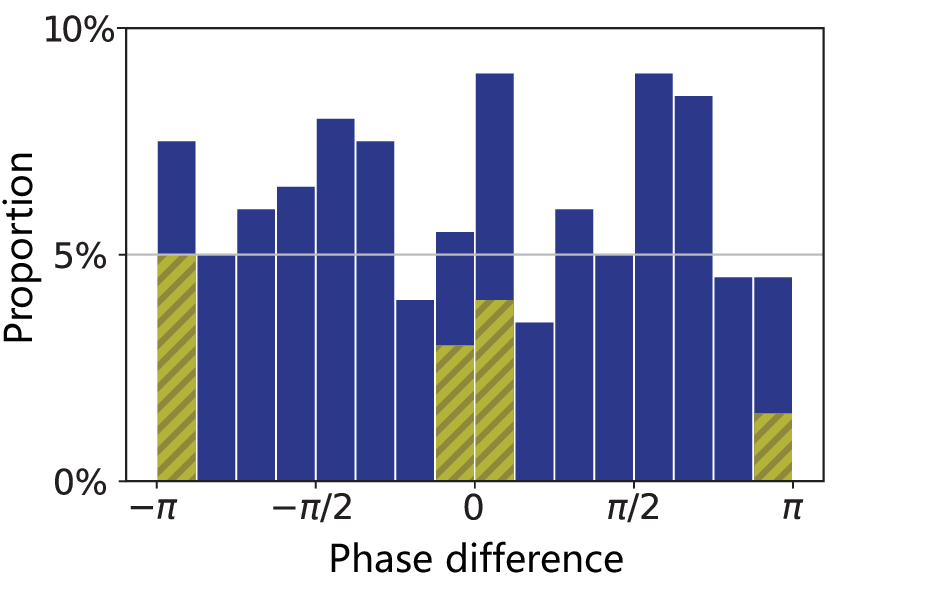}
                    \caption{
                        The distribution of the phase differences $\bm{a}^{(0)} - \bm{a}^{(1)}$ is shown.
                        The horizontal axis indicates the phase difference.
                        For a bar whose base side starts from $a$ and ends at $b$ on the horizontal axis, its height represents the proportion of DTQW-QNNs whose phase differences is between $a$ and $b$.
                        The blue and yellow striped bars together correspond to all of the DTQW-QNNs.
                        The yellow striped bars represent the portion of DTQW-QNNs whose distance after training is larger than the threshold $10^{-1}$.
                    }
                    \label{fig:phase not works}
                \end{figure}

            By comparing Figs.~\ref{fig:fixed random phases} and \ref{fig:fixed rotation axes}, we can notice that the DTQW-QNN in this section needs much more time to train compared with the DTQW-QNN in Sec.~\ref{sec:fix rand phase}.
            To reveal the cause, we have trained 200 DTQW-QNNs to approximate 200 randomly sampled operators $\hat{V}$, respectively.
            We choose a threshold to be $10^{-1}$ and mark the DTQW-QNNs of which the distance after 200 iterations of training is still larger than the threshold.
            We find that the phase differences $\bm{a}^{(0)} - \bm{a}^{(1)}$ of these marked DTQW-QNNs are all near $0$ or $\pm\pi$ as shown in Fig.~\ref{fig:phase not works}.
            Hence, we conclude that these specific differences in phases cause the DTQW-QNN to be slow to train.
            This result also corroborates that
                the phases of single-site coin operators contribute to the DTQW total effect $\hat{U}_{T,0}$ non-trivially as we have stated in Sec.~\ref{sec:fix rand phase}.
            Now knowing the cause, we can easily avoid these specific phase differences when implementing DTQW-QNNs.

        \subsection{Noise on rotation axes}
\begin{figure}[tbp]
                    \centering \includegraphics[width=\columnwidth]{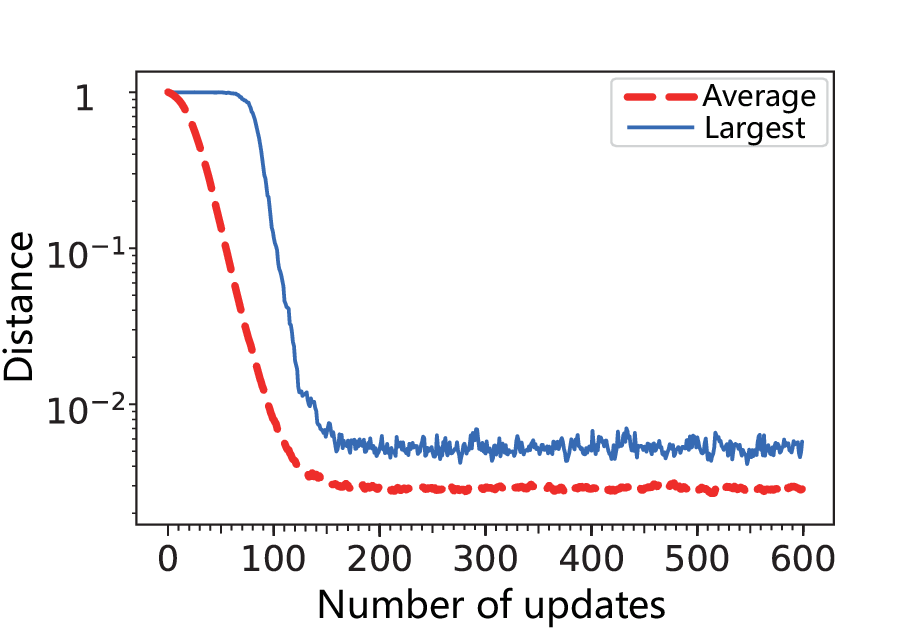}
                    \caption{\label{fig:noise on coin operators}
                        The evolution of the distance $d(\hat{U}_{T,0},\hat{V})$ during the training for a DTQW-QNN with noises present.
                        The horizontal axis represents the number of times that the DTQW-QNN is updated.
                        The vertical axis represents the distance.
                        The thick dashed red line shows the average of distances calculated from 100 sampled operators $\hat{V}$ and their corresponding DTQW-QNNs, while the thin blue line shows the largest distance among those samples.
                    }
                \end{figure}

            When the DTQW-QNN is implemented in laboratories,
                it is impossible to have all the rotation axes
                of $\hat{c}_x^{(t)}$
                be perfectly along the $x$ direction.
            There are always noises on the rotational axis, i.e.,
            \begin{equation}
                \hat{c}_{x}^{(t)} =
                        e^{i\bm{a}^{(x)}}
                        e^{i \alpha^{(x,t)} (\hat{\bm{n}}^{(x,t)}\cdot\vec\sigma)} ,
            \end{equation}
                where
                \begin{equation}
                    \hat{\bm{n}}^{(x,t)} = \begin{bmatrix}
                                0\\
                                \cos{\bm{\theta}^{(x,t)}}\\
                                \sin{\bm{\theta}^{(x,t)}}\cos{\bm{\varphi}^{(x,t)}}\\
                                \sin{\bm{\theta}^{(x,t)}}\sin{\bm{\varphi}^{(x,t)}}
                            \end{bmatrix},
                \end{equation}
                where $\bm{\theta}^{(x,t)}$ and $\bm{\varphi}^{(x,t)}$ are independent real random variables.
            In this situation,
                approximations to desired operators still can be found via gradient descent,
                as shown in Fig.~\ref{fig:noise on coin operators}.

\section{Conclusion}\label{sec:conclu}

    In conclusion, we have proposed a quantum neural network based on a simple DTQW on a cycle graph, and used the network to implement arbitrary quantum computation tasks, i.e., unitary operations on an arbitrary $N$-dimensional Hilbert space.

    In order to implement an arbitrary unitary operation via a circuit model, one needs to decompose the unitary into a sequence of smaller unitary operators. However, via our DTQW-QNN, we only need to update the parameters by a learning algorithm.
    In other words, our model is adaptive to new tasks. With a new computational task given, our network can simply evolve according to the learning algorithm, and there is no need to decompose the desired operation into a sequence of smaller gates.

    Regarding the universality of our model, we presented a specific construction of realizing arbitrary two-level unitary operations on the computational basis, and
    proved that the DTQW-QNN is universal for all unitary operations on the overall Hilbert space of the involved quantum systems.
    The DTQW-QNN is not only universal for quantum computation but also universal for controlling the whole quantum system.
    We also provided an optimization so that the circuit depth of the DTQW-QNN does not need to exceed $2n^2-2n+1$ to realize an arbitrary unitary operator on a $2n$-dimensional Hilbert space.
    However, this is only a theoretical limit of the network size in the worst case for the purpose of analytical proof. The appropriate number of nodes for each task may vary, and it is an open question to find this number for a given task.

    Our network evolves according to a learning algorithm based on gradient descent, with the loss function carefully chosen so that the parameter updates can be efficiently calculated in a back-propagation fashion and can be, in principle, directly read out from a measurement.
    The algorithm performs well in updating the parameters of the neural network. We have shown good approximations of unitary operations on a Hilbert space up to $40$ dimensions, as well as arbitrary two-outcome POVMs.
    Finally, we have also simplified the DTQW-QNN in various aspects.
    For example, the rotation gates involved in the DTQW are all limited to be along the $x$-axis. Such simplifications make the DTQW-QNN more friendly for laboratory implementations while its capability of implementing desired operations is maintained.

    We have shown the capability of the DTQW-QNNs in both analytical and numerical ways. Further studies might reveal their total capacity in completing various quantum computation tasks as well as solving machine learning problems, and further experimental implementations would make them more practically useful and closer to real-life applications.

\begin{acknowledgments}
     This work is supported by the Innovation Program for Quantum Science and Technology (Grant No. 2021ZD0301701) and the National Natural Science Foundation of China (Grant No. 12175104). Part of the numerical simulations in this work involves the use of QuTiP \cite{johansson2013qutip}.
\end{acknowledgments}


\vskip 2cm

\appendix

\section{\label{app:verif}
    Proof of Theorem~\ref{thm:universality}
}

    \begin{proof}[Proof of Theorem~\ref{thm:universality}]
        Since every unitary operator can be decomposed into
            a product of two-level unitary operators
            \cite{Nielsen2007Quantum},
            we only need to show that Theorem~\ref{thm:universality}
            stands for $\hat{V}$ of the form
            \begin{equation}\label{eq:2-level V}
                \sum_{i,j=0}^{1} v_{i,j}\ketbra{c_i,x_i}{c_j,x_j}
                +
                \sum_{
                    \substack{
                        e\neq(c_0,x_0)\\
                        e\neq(c_1,x_1)
                    }
                } \ketbra{e}{e}
                .
            \end{equation}
        We prove this by constructing
            the family of single-site coin operators
            $\{\hat{c}_x^{(t)}\}$ explicitly.

        If $c_0 \neq c_1$,
            let $t_{\mbox{meet}}$ be the solution to the integer $t$ in
            \begin{equation}\label{eq:sub11}
                \begin{cases}
                    x_0 + t\delta_{c_0} = x_1 + t\delta_{c_1} \Mod{n} \\
                    0 \leq t < n
                \end{cases}
            \end{equation}
            and $x_{\mbox{meet}}$ be
            $x_0 + t_{\mbox{meet}}\delta_{c_0} \Mod{n}$.
        The solution $t_{\mbox{meet}}$ exists and is unique since
            $\delta_0 \neq \delta_1$ and $\gcd(|\delta_0 - \delta_1|, n)=1$.
        Choose $T = n$ and
            \begin{equation}\label{eq:sub22}
                \hat{c}_x^{(t)} =
                \begin{cases}
                    \sum_{i,j=0}^1 v_{i,j}\ketsysbra{c_i}{c}{c_j} & \mbox{if $t=t_{\mbox{meet}}$ and $x=x_{\mbox{meet}}$} \\
                    \hat{I}_c & \mbox{otherwise}
                \end{cases}.
            \end{equation}
        We can verify that this $T$-step quantum walk realizes
            the two-level unitary operator $\hat{V}$ by the following calculation
    \begin{widetext}
        \begin{align}
            \hat{U}_{T,0} \sket{c,x} =& \hat{U}_{T,t_m+1}\hat{U}_{t_m+1,t_m}\hat{U}_{t_m,0} \sket{c,x} \\
            =& \hat{U}_{T,t_m+1}\hat{U}_{t_m+1,t_m} \sket{c,x+t_m\delta_c \Mod{n}} \\
            =& \begin{cases}
                \hat{U}_{T,t_m+1} \sum_{i=0}^1 v_{i,0} \sket{c_i,x_m+\delta_{c_i} \Mod{n}} & \mbox{if $(c,x) = (c_0,x_0)$} \\
                \hat{U}_{T,t_m+1} \sum_{i=0}^1 v_{i,1} \sket{c_i,x_m+\delta_{c_i} \Mod{n}} & \mbox{if $(c,x) = (c_1,x_1)$} \\
                \hat{U}_{T,t_m+1} \sket{c,x+(t_m+1)\delta_c \Mod{n}} & \mbox{otherwise}
            \end{cases} \\
            =& \begin{cases}
                \sum_{i=0}^1 v_{i,0} \sket{c_i,x_m+(n-t_m)\delta_{c_i} \Mod{n}} & \mbox{if $(c,x) = (c_0,x_0)$} \\
                \sum_{i=0}^1 v_{i,1} \sket{c_i,x_m+(n-t_m)\delta_{c_i} \Mod{n}} & \mbox{if $(c,x) = (c_1,x_1)$} \\
                \sket{c,x+n\delta_c \Mod{n}} & \mbox{otherwise}
            \end{cases} \\
            =& \begin{cases}
                \sum_{i=0}^1 v_{i,0} \sket{c_i,x_i+n\delta_{c_i} \Mod{n}} & \mbox{if $(c,x) = (c_0,x_0)$} \\
                \sum_{i=0}^1 v_{i,1} \sket{c_i,x_i+n\delta_{c_i} \Mod{n}} & \mbox{if $(c,x) = (c_1,x_1)$} \\
                \sket{c,x} & \mbox{otherwise}
            \end{cases} \\
            =& \begin{cases}
                \sum_{i=0}^1 v_{i,0} \sket{c_i,x_i} & \mbox{if $(c,x) = (c_0,x_0)$} \\
                \sum_{i=0}^1 v_{i,1} \sket{c_i,x_i} & \mbox{if $(c,x) = (c_1,x_1)$} \\
                \sket{c,x} & \mbox{otherwise}
            \end{cases}
        \end{align},
    \end{widetext}
    where $\hat{U}_{t_1,t_0}$ stands for $\mathcal{T}\prod_{t=t_0}^{t_1-1}\hat{S}\hat{C}^{(t)}$ and $t_m$, $x_m$ stands for $t_{\mbox{meet}}$ and $x_{\mbox{meet}}$ respectively.


        If $c_0 = c_1$,
            let $t_{\mbox{meet}}$ be the unique solution to the integer $t$ in
            \begin{equation}\label{eq:sub1}
                \begin{cases}
                    x_0 + t\delta_{\tilde{c}_0} = x_1 + t\delta_{\tilde{c}_1} \Mod{n} \\
                    0 < t < n
                \end{cases},
            \end{equation}
            where $\tilde{c}_0 = c_0$ and $\tilde{c}_1 = 1-c_1$.
        Denote $x_0 + t_{\mbox{meet}}\delta_{\tilde{c}_0} \Mod{n}$
            as $x_{\mbox{meet}}$.
        Choose $T = 2n$ and
            \begin{equation}\label{eq:sub2}
                \hat{c}_x^{(t)} =
                \begin{cases}
                    \hat\sigma_x & \mbox{if $t=0$ or $n$, and $x=x_1$} \\
                    \sum_{i,j=0}^1 v_{i,j}\ketsysbra{\tilde{c}_i}{c}{\tilde{c}_j} & \mbox{if $t=t_{\mbox{meet}}$ and $x=x_{\mbox{meet}}$} \\
                    \hat{I}_c & \mbox{otherwise}
                \end{cases}.
            \end{equation}
        It is easy to verify that this is a realization of
            the two-level unitary operator $\hat{V}$.
    \end{proof}

\section{Implementing the Fourier transformation\label{app:fourier}}
    In this section, we demonstrate the calculation to implement the four-by-four Fourier transformation. Firstly, we decompose the Fourier transformation $\mathrm{QFT} = \hat{u}_6\hat{u}_5\hat{u}_3\hat{u}_3\hat{u}_2\hat{u}_1$ \cite{Nielsen2007Quantum}, where
    \begin{equation}
        \hat{u}_1 = \frac{1}{2}\begin{bmatrix}
            2&0&0&0\\
            0&2&0&0\\
            0&0&\sqrt{2}&\sqrt{2}\\
            0&0&-\sqrt{2}i&\sqrt{2}
        \end{bmatrix},
    \end{equation}
    \begin{equation}
        \hat{u}_2 = \frac{1}{3}\begin{bmatrix}
            3&0&0&0\\
            0&-\sqrt{3}&-\sqrt{6}&0\\
            0&\sqrt{6}&-\sqrt{3}&0\\
            0&0&0&3
        \end{bmatrix},
    \end{equation}
        \begin{equation}
        \hat{u}_3 = \frac{1}{4}\begin{bmatrix}
            4&0&0&0\\
            0&4&0&0\\
            0&0&-1+3i&\sqrt{3}(i-1)\\
            0&0&\sqrt{3}(i+1)&-1-3i
        \end{bmatrix},
    \end{equation}
        \begin{equation}
        \hat{u}_4 = \frac{1}{2}\begin{bmatrix}
            1&-\sqrt{3}&0&0\\
            \sqrt{3}&1&0&0\\
            0&0&2&0\\
            0&0&0&2
        \end{bmatrix},
    \end{equation}
        \begin{equation}
        \hat{u}_5 = \frac{1}{3}\begin{bmatrix}
            3&0&0&0\\
            0&\sqrt{3}&-\sqrt{6}&0\\
            0&\sqrt{6}&\sqrt{3}&0\\
            0&0&0&3
        \end{bmatrix},
    \end{equation}
        \begin{equation}
        \hat{u}_6 = \frac{1}{2}\begin{bmatrix}
            2&0&0&0\\
            0&2&0&0\\
            0&0&\sqrt{2}&-\sqrt{2}\\
            0&0&\sqrt{2}&\sqrt{2}
        \end{bmatrix}.
    \end{equation}
    All these $\hat{u}_i$ are two-level unitary operators.
    By comparing $\hat{u}_i$ with Eq.~(\ref{eq:2-level V}) we can find $c_0, c_1, x_0, x_1$ for each $\hat{u}_i$.
    Then we substitute $c_0, c_1, x_0, x_1$ with their value in Eqs.~(\ref{eq:sub11}) and (\ref{eq:sub22}) if $c_0 = c_1$ or Eqs.~(\ref{eq:sub1}) and (\ref{eq:sub2}) if $c_0 \neq c_1$ to find out the DTQW for implementing each $\hat{u}_i$.
    The DTQW for each $\hat{u}_i$ is combined one after another in the temporal order of $\hat{u}_i$ to form a large DTQW.
    In other words, the walker first walks according to the DTQW for implementing $\hat{u}_1$. After the DTQW for implementing $\hat{u}_1$ is finished, the walker continues to walk according to the DTQW for implementing $\hat{u}_2$, then $\hat{u}_3$, $\hat{u}_4$, etc.
    The single-site coin operators $\hat{c}_x^{(t)}$ of the final combined DTQW for implementing the quantum Fourier transformation are shown in the following table, where X stands for the Pauli $x$ matrix and I stands for the identity matrix.
    \begin{widetext}\begin{center}
        \begin{tabular} {| c |c | c | c | c | c | c | c | c | c | c |}
        \hline
        \diagbox{$x$}{$\hat{c}_x^{(t)}$}{$t$} & 0 & 1 & 2 & 3 & 4 & 5 & 6 & 7 & 8 & 9 \\
        \hline
        0 & I & I & I & I & I & I & I & I & I & I \\
        \hline
        1 & X & $\frac{\sqrt{2}}{2}\begin{pmatrix}
            1 & -i \\
            1 & 1
        \end{pmatrix}$ & X & I & I & $-\frac{\sqrt{3}}{3}\begin{pmatrix}
            1 & \sqrt{2} \\
            -\sqrt{2} & 1
        \end{pmatrix}$ & X & $-\frac{1}{4}\begin{pmatrix}
            1+3i & \sqrt{3}(1+i) \\
            \sqrt{3}(1-i) & 1-3i
        \end{pmatrix}$ & X & I \\
        \hline\hline
        \diagbox{$x$}{$\hat{c}_x^{(t)}$}{$t$} & 10 & 11 & 12 & 13 & 14 & 15 & 16 & 17 & 18 & 19 \\
        \hline
        0 & I & $\frac{1}{2}\begin{pmatrix}
            1 & -\sqrt{3} \\
            \sqrt{3} & 1
        \end{pmatrix}$ & I & I & I & I & I & I & I & I \\
        \hline
        1 & X & I & X & I & I & $\frac{\sqrt{3}}{3}\begin{pmatrix}
            1 & -\sqrt{2} \\
            \sqrt{2} & 1
        \end{pmatrix}$ & X & $\frac{\sqrt{2}}{2}\begin{pmatrix}
            1 & 1 \\
            -1 & 1
        \end{pmatrix}$ & X & I \\
        \hline
        \end{tabular}
    \end{center}\end{widetext}

\section{\label{app:optim}
    Optimization of depth required
}
    We show in this section that any unitary operator $\hat{V}\in\mathrm{U}(2n)$ can be realized with a DTQW-based neural network of depth $2n^2-2n+1$ by constructing the implementation.

    Before the actual construction, we first introduce the follow lemma so that the total effect of our DTQW-based neural networks becomes more distinct.

    \begin{lmm}\label{lmm:totaleffect}
        For any $\hat{V}\in\mathrm{U}(2n)$,
            it is realizable by a $T$-step DTQW on an $n$-cycle
            if and only if
            \begin{equation}
                \left[
                    \mathcal{T}\prod_{\tau=0}^{T-1} \left(
                        \prod_{\xi=0}^{n-1}
                            \hat{U}^{(\xi+\tau\delta_0, \tau)}_{
                                \myket{0,\xi}, \myket{1,\xi+\tau\delta}
                            }
                    \right)
                \right] \hat{V}^\dagger \hat{S}^T = \hat{I}
            \end{equation}
            for a family of two-level unitary operators
            $\left\{\hat{U}^{(\xi, \tau)}_{
                \myket{0,\xi}, \myket{1,\xi+\tau\delta}
            }\right\}$
            indexed by the set $\{(\xi,\tau):
                0\leq \xi<n \mbox{ and } 0\leq \tau<T\}$,
            where $\hat{U}^{(\xi, \tau)}_{
                \myket{0,\xi}, \myket{1,\xi+\tau\delta}
            }$ is a two-level unitary acting on the subspace spanned by
            $\{\myket{0,\xi}, \myket{1,\xi+\tau\delta}\}$,
            and $\delta = \delta_0 - \delta_1$.
    \end{lmm}

    This lemma is proved by the following calculation:
        \begin{eqnarray}
            \hat{U}_{T, 0} &=& \mathcal{T} \prod_{t=0}^{T-1}
                \hat{S} \hat{C}^{(t)},
            \\
            \hat{U}_{T, 0} &=& \mathcal{T} \prod_{t=0}^{T-1} \Bigg[
                \hat{S} \cdot \prod_{x=0}^{n-1} \Bigg(
                    \begin{aligned}[t]
                        & \hat{c}_{x}^{(t)} \otimes \ketbra{x}{x}
                        \\
                        & + \sum_{\xi \neq x}
                            \hat{I}_{x} \otimes \ketbra{\xi}{\xi}
                \Bigg)
            \Bigg],
                    \end{aligned}
              \\
           \hat{U}_{T, 0} &=&\mathcal{T}\prod_{t=0}^{T-1}\Bigg[
                \begin{aligned}[t]
                &   \hat{S} \cdot \hat{S}^{t} \cdot                             \hat{S}^{-t}\prod_{x=0}^{n-1}\Bigg(
	                    \hat{c}_{x}^{(t)} \otimes|x\rangle\langle x|
	         \\
        	            &+ \sum_{\xi \neq x} \hat{I}_{x} \otimes| \xi\rangle\langle\xi|\Bigg) \hat{S}^{t} \cdot \hat{S}^{-t}\Bigg],
                \end{aligned}
        \end{eqnarray}
        \begin{eqnarray}
            \hat{U}_{T, 0}
           &=& \mathcal{T} \prod_{t=0}^{T-1}\Bigg[\begin{aligned}[t]&\hat{S}^{t+1} \prod_{x=0}^{n-1}            \Bigg(
	                    \hat{S}^{-t} \cdot \hat{c}_{x}^{(t)} \otimes|x\rangle\langle x|
	                 \\
                    &+\sum_{\xi \neq x} \hat{I}_{x} \otimes| \xi\rangle\langle\xi| \cdot \hat{S}^{t}\Bigg) \cdot \hat{S}^{-t}\Bigg],
                     \end{aligned}
              \\
         \hat{U}_{T, 0} &=&S^{T} \cdot \mathcal{T} \prod_{t=0}^{T-1}\left(\prod_{x=0}^{n-1} \hat{U}^{(x, t)}\right).
        \end{eqnarray}
        Notice that if $\xi+t \cdot \delta_{c} \neq x$,
        \begin{equation}
        \hat{U}^{(x, t)}|c, \xi\rangle=|c, \xi\rangle.
        \end{equation}
        Hence, $\hat{U}^{(x, t)}$ is a two-level unitary, and the possible nonidentity effect subspace is spanned by
        $\left\{\left|0, x-t \delta_{0}\right\rangle,\left|1, x-t \delta_{1}\right\rangle\right\}$. Thus
        \begin{eqnarray}
        \hat{U}_{T, 0}&=&S^{T} \cdot \mathcal{T} \prod_{t=0}^{T-1}\left(\prod_{x=0}^{n-1} \hat{U}_{\left|0, x-t \delta_{0}\right\rangle,\left|1, x-t \delta_{1}\right\rangle}^{(x, t)}\right)\nonumber
        \\
        &=&S^{T} \cdot \mathcal{T} \prod_{t=0}^{T-1}\left(\prod_{\xi=0}^{n-1} \hat{U}_{|0, \xi\rangle,|1, \xi+t \delta\rangle}^{\left(\xi+t \delta_{0, t}\right)}\right) ,
        \end{eqnarray}
        where $\delta=\delta_{0}-\delta_{1}, \xi=x-t \delta_{0}$.
        By moving all shift operators in Eq.~(\ref{eq:DTQW U}) to the far left, this lemma is proved.

    With this lemma, we can finally start our construction of the implementation for arbitrary unitary operators $\hat{V}$. Let us denote
    \begin{equation}
    \hat{V}_{t}=
    \left\{
    \begin{array}{ll}
    \hat{V}_{t} \cdot \hat{V}_{t-1} & \mbox{if $t \geq 2$} \\
    \prod_{\xi=0}^{n-1} \hat{U}_{|0, \xi\rangle,|1, \xi+t \delta\rangle}^{\left(\xi+t \delta_{0, t}\right)}
    \cdot \hat{S}^{-1} & \mbox{if $t =1$}; \\
    \hat{V}_{0} \cdot \hat{S}^{-1} & \mbox{if $t =0$}
    \end{array}
    \right.
    \end{equation}

    if $2kn \leqslant \tau<2kn+n-k$ and $\xi=(k-1) \delta \Mod{n}$:
        \begin{equation}
        \hat{U}^{\left(\xi+\tau \delta_{0}, \tau\right)}=
        \hat{U}_{x|1, \xi+\tau \delta\rangle}^{\left(\xi+\tau \delta_{0}, \tau\right)}
        \left(\tilde{V}_{\tau}|0, k \delta\rangle\right),
        \end{equation}

    if $2kn \leqslant \tau<2kn+n-k-1 $ and $\xi=-\delta\Mod{n}$:
        \begin{equation}
        \hat{U}^{\left(\xi+\tau \delta_{0}, \tau\right)}=
        \hat{U}_{x|0, \xi\rangle}^{\left(\xi+\tau \delta_{0}, \tau\right)}\left(\tilde{V}_{\tau}|0, k \delta\rangle\right),
        \end{equation}

    if $2kn+n+k \leqslant \tau<2(k+1)n$ and
    $\xi=(k-1-t)\delta\Mod{n}$:
        \begin{equation}
        \hat{U}^{\left(\xi+\tau \delta_{0}, \tau\right)}=
        \hat{U}_{x|0, \xi\rangle}^{\left(\xi+\tau \delta_{0}, \tau\right)}\left(\tilde{V}_{\tau}|1, k \delta\rangle\right),
        \end{equation}

    if $(2k+1)n \leqslant \tau <2(k+1)n-k$ and $\xi=k\delta\Mod{n}$:
        \begin{equation}
        \hat{U}^{\left(\xi+\tau \delta_{0}, \tau\right)}=
        \hat{U}_{x|1, \xi+\tau \delta\rangle}^{\left(\xi+\tau \delta_{0}, \tau\right)}\left(\tilde{V}_{\tau}|1, k \delta\rangle\right),
        \end{equation}
     where $\delta=\delta_{0}-\delta_{1}$, $k=\floor{\frac{\tau}{2 n}}$, and $U_{x|\varphi\rangle}^{\left(\xi+\tau \delta_{0}, \tau\right)}(|\psi\rangle)$ is any two-level unitary subject to $\left\langle\varphi\left|U_{x|\varphi\rangle}^{\left(\xi+\tau \delta_{0}, \tau\right)}\right| \psi\right\rangle= 0$.
    One can easily verify that such $U_{x|\varphi\rangle}^{\left(\xi+\tau \delta_{0}, \tau\right)}(|\psi\rangle)$ always exists as long as  $|\psi\rangle=|0, \xi\rangle$ or $|1, \xi+\tau \delta\rangle$.

    For induction on $\floor{\frac{t}{2 n}}$, let $t=2n^{2}-2n+1$,
    With $c_{x}^{(t)}=\tensor[_p]{\mybra{x}}{}\hat{S}^{t} \hat{U}^{(x, t)} \hat{S}^{-t}\sket{x}_p$, $\forall c \leqslant 1$, $\forall l<\floor{\frac{t}{2 n}}$, we have
    \begin{equation}
    \hat{V}_{t}|c, l \delta\rangle=|c, l\delta\rangle.
    \end{equation}



\section{Controlling large systems via DTQW-based neural network\label{app:ctrl large sys}
}
        \begin{figure}[htbp]
            \centering
                \includegraphics[width= 0.9\columnwidth]{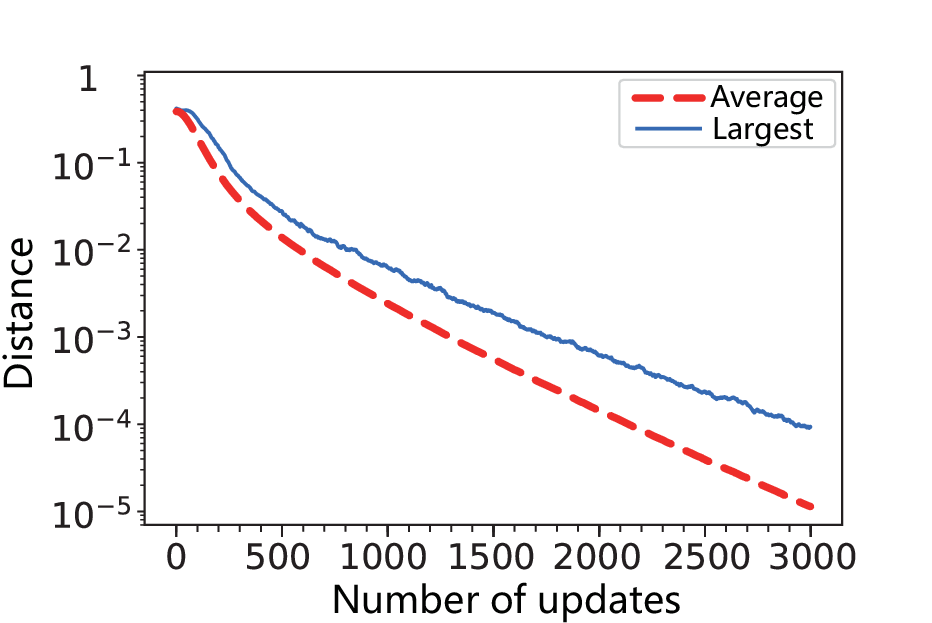}
            \caption{
                The evolution of the distance between the DTQW and the desired operation during the training of DTQW-QNNs.
                The horizontal axis represents the number of times that the DTQW-QNN is updated. The vertical axis represents the largeness of the distance.
                The thick dashed red line shows the average of distances calculated from 200 sampled operations and their corresponding DTQW-QNNs, while the thin blue line shows the largest distance among those samples.
            }\label{fig:indirect control unitary}
        \end{figure}

        In Sec.~\ref{sec:gradient}, we mentioned the possibility of
            controlling a large system via the DTQW indirectly
            by controlling the $2$-level coin system.
        As shown in Fig.~\ref{fig:indirect control unitary},
            this is actually feasible, indicated by the numerical results,
            when the desired operation on the position system is unitary.

        \begin{figure}[htbp]
                \centering
                \includegraphics[width= 0.9\columnwidth]{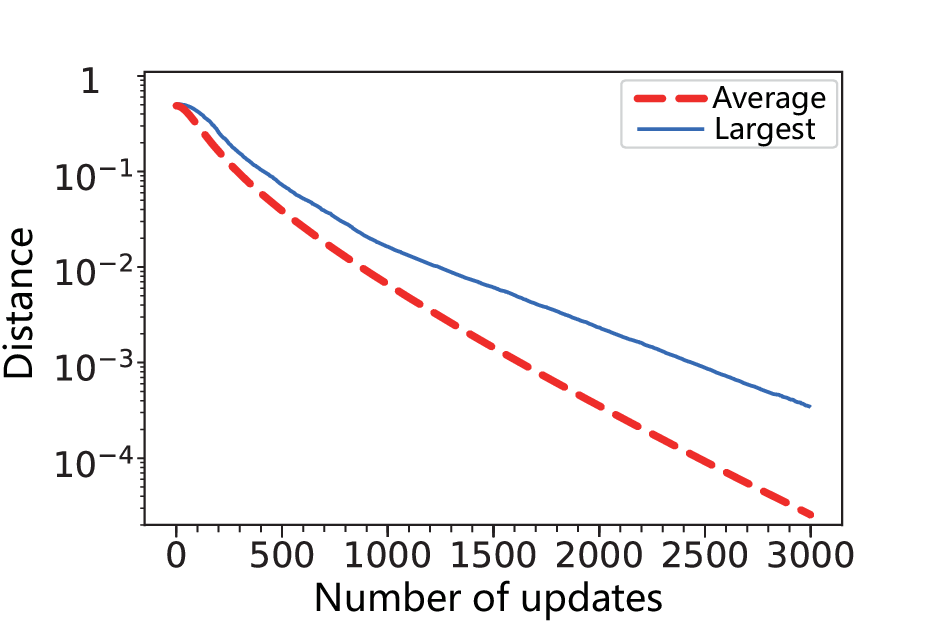}
            \caption{
                The evolution of the distance during the training of DTQW-QNNs.
                The horizontal axis represents the number of times that the DTQW-QNN is updated. The vertical axis represents the distance.
                The thick dashed red line shows the average of distances calculated from 200 sampled operations and their corresponding DTQW-QNNs, while the thin blue line shows the largest distance among those samples.
            }\label{fig:indirect control POVM}
        \end{figure}

        Not only unitary operations can be realized in this indirect controlling fashion,
            but more general quantum operations such as POVM measurements can also be realized, as shown in Fig.~\ref{fig:indirect control POVM}.
        To apply gradient descent in this situation,
            the loss is defined as
            \begin{equation}
                L_{\sket{\psi}_p} = \frac{1}{2} \sum_{j=0}^1 \left\lVert
                        \sket{\psi^{(T)}_j}_p - \sket{\phi^{(T)}_j}_p
                    \right\rVert^2 ,
            \end{equation}
            where $\sket{\psi^{(T)}_j}_p =
                \tensor[_c]{\mybra{j}}{} \hat{U}_{t,0}\sket{0}_c \sket{\psi}_p$,
            and $\sket{\phi^{(T)}_j}_p = \hat{M}_j \sket{\psi}_p$.
        This loss is well-selected by us so that
            the form of partial derivatives in Eq.~(\ref{eq:partial}) needs no modification, i.e.,
            \begin{equation}\label{eq:partial povm}
                \prtl{L_{\sket{\Psi}}}{\alpha^{(x,t)}_j} =
                    \operatorname{Im}\left(
                        \sbraopket{\Phi^{(t)}}{
                            \hat\Sigma_j^{(x,t)}
                        }{\Psi^{(t)}}
                    \right),
            \end{equation}
            where $\sket{\Psi^{(t)}} = \hat{U}_{t,0}\sket{0}_c\sket{\psi}_p$,
            $\sket{\Phi^{(t)}} = \hat{U}_{t,0}^\dagger\sket{\Phi^{(T)}}$,
            and $\sket{\Phi^{(T)}} = \sum_{j=0}^1 \sket{j}_c\sket{\phi^{(T)}_j}_p$.
        The distance between two measurements
            $\{\hat{N}_j=\tensor[_c]{\mybra{j}}{} \hat{U}_{t,0}\sket{0}_c\}_{j=0}^1$ and $\{\hat{M}_j\}_{j=0}^1$ in Fig.~\ref{fig:indirect control POVM} is measured by
            \begin{multline}
                d(\{\hat{N}_j\}_{j=0}^1,\{\hat{M}_j\}_{j=0}^1)
                    \\=
                        \frac{1}{2n\sqrt{2}} \sum_{j=0}^1 \sqrt{
                            \mathrm{tr}^2(\hat{M}_j^\dagger\hat{M}_j)
                            +\mathrm{tr}^2(\hat{N}_j^\dagger\hat{N}_j)
                            -2|\mathrm{tr}(\hat{N}_j^\dagger\hat{M}_j)|^2
                        } .
            \end{multline}

\section{Meta parameters used in numerical simulation}\label{app:metapara}
    For all numerical simulations, the $\delta_0$ and $\delta_1$ for the shift operator $\hat{S}$ are set to be $0$ and $1$ respectively.
    And all the real initial parameters $\alpha^{(x,t)}_j$ in the coin operators before the training are randomly sampled from $[-2\pi, 2\pi]$ uniformly and independently.
    The training sets are always the Haar-measured pure states from the appropriate Hilbert space. For the desired operator $\hat{V}$, the number of depth $T$, the number of sites in the cycle $n$, the learning rate $\eta$, the number of samples of DTQW-QNN trained in parallel $N_\text{sample}$ and other randomness involved, see the table below [where $\mathrm{U}(2)$ and $\mathrm{U}(4)$ are equipped with corresponding Haar measures].

    \begin{widetext}
        \begin{center}
        \begin{tabular}{||c | c | c | c | c | c | c||}
         \hline
         Figure & $\hat{V}$ & $T$ & $n$ & $\eta$ & $N_\text{sample}$ & Other randomness\\ [0.5ex]
         \hline\hline
         Fig.~\ref{fig:SWAP} & $\mathrm{SWAP}$ & 5 & 2 & 0.1 & / & /\\
         \hline
         Fig.~\ref{fig:dist distri V total} & $\mathrm{U}(2)$ & / & 2 &0.05 &200 for each T&/\\
         \hline
         Fig.~\ref{fig:dist drop V total} & $\mathrm{QFT}$ & $2n^2$ & / &0.05 &\begin{tabular}{@{}c@{}}200 for $n=2,3$ \\ 50 for $n=4,5$\end{tabular} &/\\
         \hline
         Fig.~\ref{fig:acc by overpara} & $\mathrm{QFT}$ & / & 2 &0.01 &200 for each T&/\\
         \hline
         Fig.~\ref{fig:large sys} & $\mathrm{QFT}$ & 500 & 20 &0.05 &10&/\\
         \hline
         Fig.~\ref{fig:fixed random phases} & $\mathrm{U}(4)$ & 20 & 2 &0.05 &200&$\bm{a}^{(x)}$ uniformly sampled from $[0, 2\pi]$\\
         \hline
         Fig.~\ref{fig:fixed rotation axes} & $\mathrm{U}(4)$ & 20 & 2 &0.05 &200&$\bm{a}^{(x)}$ uniformly sampled from $[0, 2\pi]$ and shared by all samples\\
         \hline
         Fig.~\ref{fig:phase not works} & $\mathrm{U}(4)$ & 20 & 2 &0.05 &200&$\bm{a}^{(x)}$ uniformly sampled from $[0, 2\pi]$ and independent for all samples\\
         \hline
         Fig.~\ref{fig:noise on coin operators} & $\mathrm{QFT}$ & 20 & 2 &0.1 &100&\begin{tabular}{@{}c@{}}$\bm{a}^{(x)}$ uniformly sampled from $[0, 2\pi]$ and shared by all samples \\ $\bm{\theta}^{(x,t)}$ sampled from normal distribution with standard derivation $0.01$ \\ $\bm{\varphi}^{(x,t)}$ uniformly sampled from $[0,2\pi]$ \end{tabular}\\
         \hline
         Fig.~\ref{fig:indirect control unitary} & $\mathrm{U}(2)$ & 20 & 4 &0.01 &150&/\\ [1ex]
         \hline
         Fig.~\ref{fig:indirect control POVM} & $\mathrm{U}(4)$ & 20 & 4 &0.01 &150&/\\ [1ex]
         \hline
        \end{tabular}
        \end{center}
    \end{widetext}

\end{document}